\documentclass[reprint,superscriptaddress,
amsmath,amssymb,aps,pra,twocolumn]{revtex4-2}
\usepackage{graphicx}
\usepackage{dcolumn}
\usepackage{newtxtext}
\usepackage{newtxmath}
\usepackage{bm}     
\usepackage[breaklinks=true,colorlinks,citecolor=blue,linkcolor=blue,urlcolor=blue]{hyperref}
\usepackage{xcolor}
\usepackage{orcidlink}

\newcommand{\tr}[1]{{\text{tr}{#1}}}

\begin{document}
%-----------------------------------------------------------
\title{Algebraic power scaling in a slowly-quenched bosonic quantum battery}
%-----------------------------------------------------------
\author{Donny~\surname{Dwiputra}\orcidlink{0000-0002-3645-7892}}\email[Corresponding author: ]{donny.dwiputra@ymail.com}
\affiliation{Research Center for Quantum Physics, National Research and Innovation Agency (BRIN), South Tangerang 15314, Indonesia}
\affiliation{Asia Pacific Center for Theoretical Physics, Pohang 37673, South Korea}
%-----------------------------------------------------------
\author{Ahmad R. T. \surname{Nugraha}\orcidlink{0000-0002-5108-1467}}\email{ahmad.ridwan.tresna.nugraha@brin.go.id}
\affiliation{Research Center for Quantum Physics, National Research and Innovation Agency (BRIN), South Tangerang 15314, Indonesia}
\affiliation{Department of Engineering Physics, School of Electrical Engineering, Telkom University, Bandung 40257, Indonesia}
%-----------------------------------------------------------
\author{Sasfan A. \surname{Wella}\orcidlink{0000-0002-5884-4287}}\email{sasfan.arman.wella@brin.go.id}
\affiliation{Research Center for Quantum Physics, National Research and Innovation Agency (BRIN), South Tangerang 15314, Indonesia}
%-----------------------------------------------------------
\author{Freddy P. \surname{Zen}\orcidlink{0009-0007-7150-3798}}\email{fpzen@fi.itb.ac.id}
\affiliation{Theoretical Physics Laboratory, Department of Physics, Faculty of Mathematics and Natural Sciences, Institut Teknologi Bandung, Bandung 40132, Indonesia}
\affiliation{Indonesian Center for Theoretical and Mathematical Physics, Bandung 40132, Indonesia}
%-----------------------------------------------------------
%-----------------------------------------------------------
\begin{abstract}
%-----------------------------------------------------------
Bosonic modes provide a promising platform for quantum batteries as a result of their unbounded energy spectrum. However, the energy that can be stored during a coherent charging process is limited due to coherent oscillations between the charger and battery. In this work, we show that by introducing a slow quench in the interaction between a coherently driven charger mode and a quadratic oscillator battery,
the maximum stored energy and maximum battery power scale algebraically with the quench duration, with exponents controlled by the ramp profile.
This finding implies that, quite counterintuitively, slower quenches lead to faster charging. Such a quench suppresses coherent energy oscillations between the battery and the charger, allowing an unbounded increase in power.
We further show that, in the ideal closed protocol, the stored energy is fully extractable as ergotropy, while charger dissipation converts the algebraic enhancement into a finite-time scaling window with an optimal quench duration.
We also show that the temporal-extensive scaling occurs in a broader context by mapping the system to a coherently driven Tavis-Cummings battery.
Finally, we discuss experimentally accessible signatures in superconducting circuit quantum electrodynamics and organic microcavity platforms.
\end{abstract}
%-----------------------------------------------------------
\date{\today}
\maketitle
%-----------------------------------------------------------
\section{Introduction}
%-----------------------------------------------------------
Quantum thermodynamics is a rapidly expanding field that seeks to understand the role of coherence and entanglement in developing thermodynamic protocols to outperform their classical counterparts~\cite{gemmer2009quantum,kosloff2013quantum,millen2016perspective,vinjanampathy2016quantum,goold2016role,binder2018thermodynamics,alicki2019introduction,campbell2026roadmap}.  To avoid speed bottlenecks in a quantum engine, it is crucial to identify efficient methods for energy storage and utilization.  This quest naturally motivates the concept of a quantum battery~\cite{alicki2013entanglement,hovhannisyan2013entanglement,binder2015operational,binder2015quantacell,campaioli2017enhancing,ferraro2018high,le2018spin,henao2018role,andolina2019extractable,barra2019dissipative,julia2020bounds,gyhm2022quantum,quach2023quantum,ahmadi2024nonreciprocal,lu2025topological,campaioli2024colloquium}, which is an interacting quantum system (comprising a charger and a battery) utilizing coherence or beneficial entanglement \cite{kamin2020entanglement,shi2022entanglement,gyhm2024beneficial} to provide significant advantages over its classical counterpart, particularly in terms of charging power. In such a system, the charger acts as a transducer that converts an energy source, e.g., an excited state or coherent driving, to a battery excitation in the charging process. How the battery maintains the stored energy after charging or the discharging process~\cite{santos2021quantum,mohan2021reverse,arjmandi2022enhancing,xu2023charging,song2025self} will, however, not be the subject of this present work.

A wide variety of quantum battery models have been introduced in the literature, for instance in Refs.~\citenum{ferraro2018high,andolina2018charger,farina2019charger,ukhtary2023high,downing2023quantum,downing2024hyperbolic,downing2024energetics,downing2021exceptional,downing2024two,downing2025energy}.  These models typically consider interacting systems where either the charger or the battery is composed of spins or bosonic oscillators.  In spin-based batteries, interactions between battery units can lead to superextensive scaling of power with the number of battery units $N$, as demonstrated in Sachdev-Ye-Kitaev~\cite{rossini2020quantum,rosa2020ultra,francica2024quantum,romero2025scrambling,divi2025sachdev} and finite Dicke~\cite{campaioli2024colloquium,ferraro2018high,crescente2020charging,crescente2020ultrafast,dou2022extended} quantum batteries. This so-called quantum power advantage represents a speed-up in optimal charging time compared to parallel charging, for which the charging power scales linearly with $N$.  However, spin batteries can only store a limited amount of energy since they have bounded energy levels, whereas oscillator energies are unbounded from above.

Despite their apparent advantage, the existing oscillator quantum battery models are suboptimal in terms of stored energy when coherent driving is applied over a given charging duration. In typical oscillator battery models~\cite{andolina2018charger,farina2019charger}, only a few lowest excited battery states can be occupied at maximum, irrespective of the charging duration.  This limitation ignores the advantage of the extensive capacity of oscillator batteries.  Some studies use time-dependent control to optimize the driving source, but this approach results in a rather complicated time dependence~\cite{mazzoncini2023optimal,rodriguez2024optimal}. Other time-dependent protocols, such as using linear feedback control in an open quantum battery \cite{mitchison2021charging}, exhibit a relatively modest improvement in the maximum stored energy.

In this work, we demonstrate that a continuous-time quench of the charger-battery interaction provides a simple mechanism for overcoming the energy limitation of coherently driven oscillator batteries. Instead of producing only bounded energy oscillations, a sufficiently slow ramp allows the coherent drive to accumulate excitation over an extended time and transfer it efficiently to the battery. As a result, both the maximum stored energy and the maximum charging power grow algebraically with the ramp duration, with exponents determined by the ramp profile. In the ideal closed system, this scaling implies that arbitrarily high oscillator excitations become accessible in the long-ramp limit. We refer to this behavior as temporal extensivity, since the enhancement is controlled by the ramp duration rather than by increasing the number of battery units. We also analyze how charger dissipation limits this algebraic regime and discuss the relation of the model to a driven Tavis-Cummings battery.

%-----------------------------------------------------------
\begin{figure}[tb]
\centering\includegraphics[clip,width=7cm]{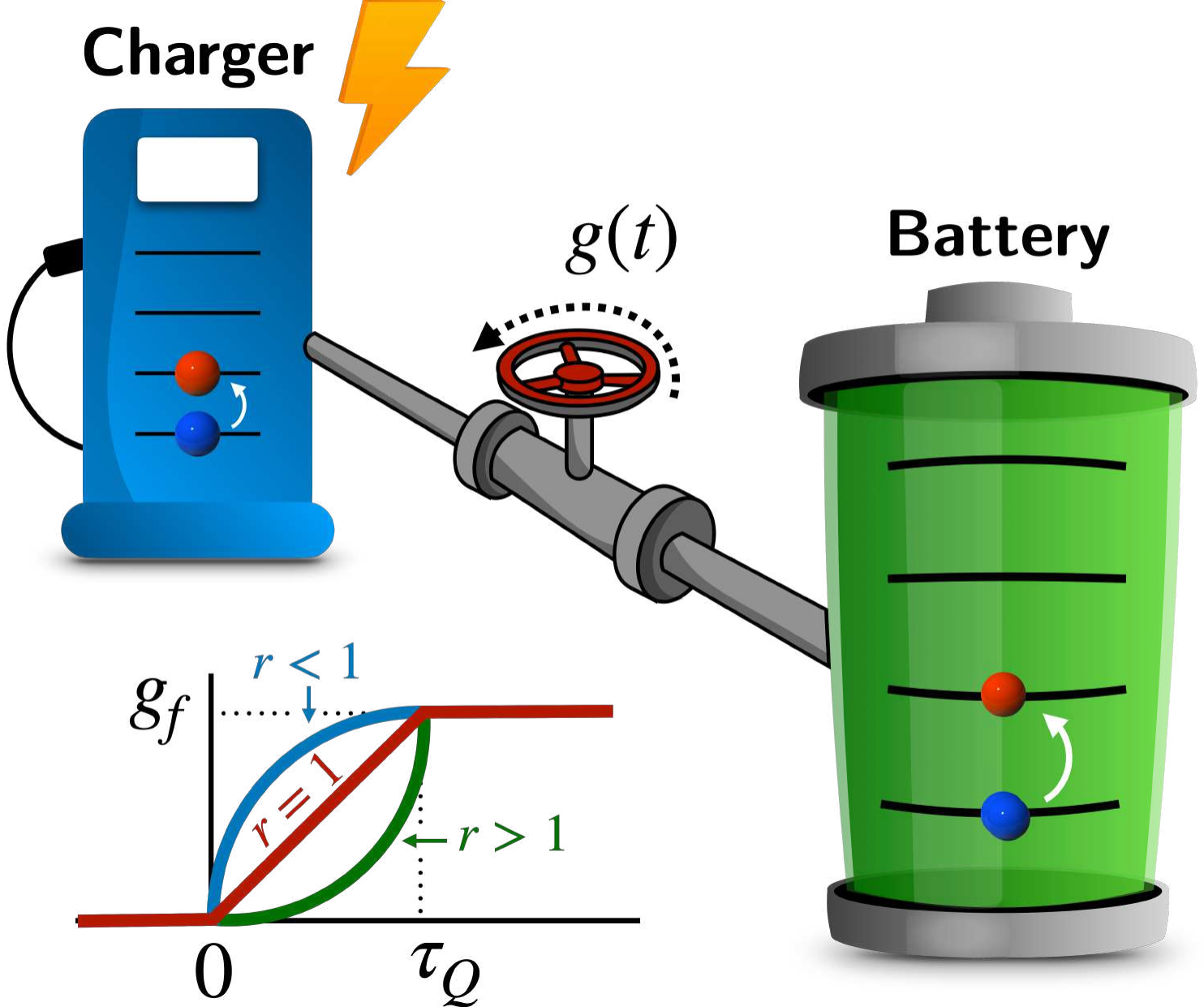}
\caption{\label{fig1} Illustration of the oscillator quantum battery with a continuously quenched coupling $g(t)$. The charger is powered with a coherent external drive. Both the battery and charger are initially at the ground state.}
\end{figure}
%----------------------------------------------------------- 

%----------------------------------------------------------- 
\section{Quantum Battery Model}
%-----------------------------------------------------------

We model the charging process between the charger $\text A$ and the battery $\text B$ in resonance with the driving as a nonequilibrium interaction within a quench duration $\tau_Q$, depicted in Fig.~\ref{fig1} for $r < 1$, $r = 1$ (linear quench), and $r > 1$. We consider the  Hamiltonian $H(t)$ (with a natural unit $\hbar=1$) in the frame rotating with the drive as follows,
\begin{equation}\label{model}
    H(t) = g(t)\left(a b^\dag + a^\dag b\right) + F(a + a^\dag)
\end{equation}
where $a, b$ ($a^\dag,b^\dag)$ are the bosonic annihilation (creation) operators of the A and B subsystems, respectively. Here, $F$ is the external driving amplitude applied to the charger only, whereas the quenched coupling interaction $g(t)$ with  maximum strength ${g_{f}}$ and exponent $r\in\mathbb{R}^+$ is defined as
\begin{equation}
g(t) = 
    \begin{cases}
       {g_{f}} (t/\tau_Q)^r, & 0\leq t\leq \tau_Q,\\
       {g_{f}}, & t > \tau_Q.
    \end{cases}
\end{equation}
To compare with the case where the coupling is turned ``on/off'' immediately, we can keep the coupling in its maximum value after $\tau_Q$. At time $t=0$, both charger and battery are in the ground state $|\psi(0)\rangle_{AB}=|0\rangle_{A}|0\rangle_{B}$, where $|0\rangle_{A,B}$ is the respective ground state of $H_{A,B}$ with $H_A=\omega_0 a^\dag a$ and $H_B=\omega_0 b^\dag b$. Note that since $[ab^\dag+a^\dag b, H_{A}+H_{B}]=0$, the only source of energy injected into the charger is external driving.

We also incorporate the environmental effect where the charger is subject to dissipation, quantified by $\gamma$.  The effective dynamics is described by the following Gorini-Kossakowski-Sudarshan-Lindblad (GKSL) master equation \cite{gorini1976completely, lindblad1976onthegenerators},
\begin{equation}\label{lindblad}
    \dot{\rho}_{AB} = \mathcal L \rho_{AB} = -i[H(t),\rho_{AB}] + \gamma\mathcal D[a]\rho_{AB},
\end{equation}
where $\mathcal D[A] \rho = A\rho A^\dag - \frac{1}{2}\{A^\dag A,\rho\}$ is the GKSL dissipator. To calculate the Heisenberg equation of motion of an operator $A(t)$, one considers the adjoint equation $\dot{A}=\mathcal L^\dag A$, by the cyclicity of trace in $\langle A(t)\rangle = \tr[A(0)e^{\mathcal L t}\rho(0)]=\tr[e^{\mathcal L^\dag t}A(0)\rho(0)]$. 
For the calculation of the first moments below, the quantum-jump term in Eq.~\eqref{lindblad} does not contribute beyond the damping term in the adjoint equation. Equivalently, the same equations of motion for $\langle a\rangle$ and $\langle b\rangle$ are obtained from the effective non-Hermitian Hamiltonian~\cite{ashida2020nonhermitian},
\begin{equation}\label{H_NH}
H_{NH}(t)=H(t)-i\frac{\gamma}{2}a^\dagger a.
\end{equation}
This equivalence is used only at the level of first moments and should not be understood as an equivalence between the full GKSL evolution and the non-Hermitian dynamics at the level of density matrices.
Compared to using the full Liouvillian in Eq.~\eqref{lindblad}, the computational time will be reduced quadratically. Both formalisms result in the same equations of motion, Eq.~\eqref{heis_eom}. 

%-----------------------------------------------------------
\section{Charging dynamics and temporal extensivity}
%-----------------------------------------------------------

To calculate the energy stored in the battery, $E_{B}(t)=\tr[H_B\rho(t)]$, and the average power, $P_{B}(t)=E_{B}(t)/t$, we first consider the equations of motion for the first moments,
\begin{equation}\label{heis_eom}
    \langle\dot{a}\rangle=-i\left(g(t)\langle b\rangle + F \right)-\frac{\gamma}{2}\langle a\rangle, \quad \langle\dot{b}\rangle=-ig(t)\langle a\rangle.
\end{equation}
We begin with the constant-coupling case, $r=0$, and a dissipationless charger, $\gamma=0$. The energy stored in the battery after a time $t$ is
\begin{equation}\label{vanilla}
    E^{(r=0)}_{B}(t)
    =
    4\frac{\omega_0 F^2}{{g_{f}}^2}
    \sin^4\!\left(\frac{{g_{f}} t}{2}\right),
\end{equation}
from which the maximum stored energy is
$E^{(r=0)}_{B,m}=4\omega_0 F^2/{g_{f}}^2$.
This result was originally derived in Ref.~\citenum{farina2019charger} and follows directly from the Heisenberg equations for the constant-coupling Hamiltonian. For completeness, we provide the derivation in Appendix~\ref{app:constant_coupling}.

We define $t_\text{m}$ as the earliest time at which $E_{B}$ reaches its first maximum. For $r=0$, this occurs at $t^{(r=0)}_\text{m}=\pi/{g_{f}}$. The corresponding maximum average power is therefore $P^{(r=0)}_{B,m}=4\omega_0 F^2/({g_{f}}\pi)$, which is independent of the quench duration $\tau_Q$. The constant coupling model $r=0$ therefore provides no charging advantage over a longer quench duration $\tau_Q$. 

Now, we turn to the continuous quench with $r>0$. We can decouple Eq.~\ref{heis_eom} for the battery part, for $t\leq\tau_Q$,
\begin{equation}\label{eom}
    t\langle\ddot{b}\rangle+\left(\frac{\gamma t}{2} - r\right)\langle\dot{b}\rangle+ k^2 t^{2r+1}\langle b\rangle = -k t^{r+1}F,
\end{equation}
where $k={g_{f}}/\tau_Q^r$. 
For a closed system, $\gamma=0$, Eq.~(\ref{eom}) becomes a linear nonhomogeneous Emden-Fowler equation~\cite{wong75onthegeneralized}.  The result can be written in terms of the coherent amplitudes
\begin{equation}
A(t)=\langle a(t)\rangle,\qquad B(t)=\langle b(t)\rangle .
\label{eq:coh-amplitudes}
\end{equation}
Since the Hamiltonian is quadratic and the initial state is the two-mode vacuum, the dynamics preserves a product of coherent states, and thus the moments factorize and $E_B(t)=\omega_0|B(t)|^2$. Solving the first-moment equations during the ramp gives the integral representation
\begin{equation}\label{E_B}
E_B(t)
=
\omega_0F^2t^2
\left[
\int_0^1
\sin\left(
\frac{\theta(t)}{1+r}
\left[1-u^{r+1}\right]
\right)du
\right]^2 ,
\end{equation}
where
\begin{equation}
\theta(t)=kt^{1+r}
=
\frac{g_f t^{1+r}}{\tau_Q^r}.
\end{equation}
This expression is equivalent to the Green-function solution of Eq.~(\ref{eom}), but it makes the scaling with $\tau_Q$ explicit. For a fast quench, $\tau_Q\lessapprox \pi {g_{f}}^{-1}$, the quench is too fast so that the first maximum of energy is not yet reached; the limiting case for $r=0$ of Eq.~(\ref{E_B}) is given by Eq.~(\ref{vanilla}). Meanwhile, for a slow enough quench, $\tau_Q\gtrapprox g_f^{-1}$, $E_B(t)$ grows quadratically until it reaches $E_{B,m}$ that scales monotonically with respect to $\tau_Q$, as demonstrated below.

The first maximum of $E_B(t)$ is reached at a dimensionless value $\theta(t_m)=\theta_m$, where $\theta_m$ depends on the ramp exponent $r$, but not on $\tau_Q$.  Since
\begin{equation}
\theta_m
=
\frac{g_f t_m^{1+r}}{\tau_Q^r},
\end{equation}
the first maximum time scales as
\begin{equation}
t_m
=
\left(\frac{\theta_m}{g_f}\right)^{1/(1+r)}
\tau_Q^{r/(1+r)} .
\end{equation}
Defining
\begin{equation}
\alpha=\frac{r}{r+1},
\end{equation}
we obtain $t_m\propto \tau_Q^\alpha$.  Evaluating Eq.~(\ref{E_B}) at $t=t_m$ then gives
\begin{equation}\label{scaling}
    E_{B,m}\propto \tau_Q^{2\alpha}, \quad
    P_{B,m} = \frac{E_{B,m}}{t_m} \propto \tau_Q^\alpha,
\end{equation}
which is the main result of this study. For $r\in[0,\infty)$, the power-law exponent is bounded by $\alpha\in[0,1]$.  The derivation of Eq.~(\ref{E_B}) and the scaling relation in Eq.~(\ref{scaling}) is given in Appendix~\ref{app:slow_quench_derivation}.
Note that the scaling in Eq.~\eqref{scaling} assumes that the first maximum occurs during the ramp interval, $t_m\leq \tau_Q$. Since $\theta(\tau_Q)=g_f\tau_Q$, this requires $\theta_m\leq g_f\tau_Q$, or equivalently $\tau_Q\geq \theta_m/g_f$.

\begin{figure}[tb]
\centering\includegraphics[width=\columnwidth]{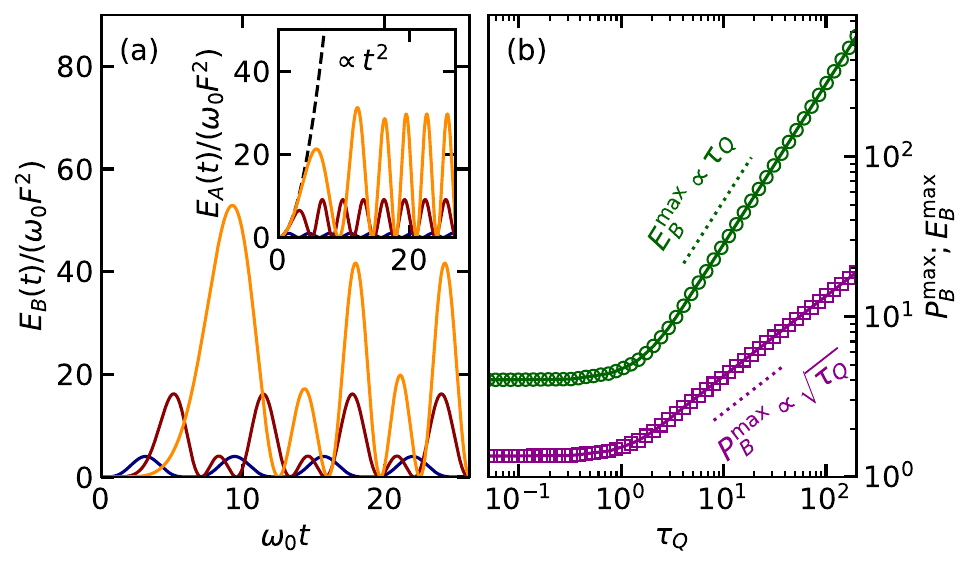}
\caption{\label{fig2} 
Temporal extensivity of the charging process in a lossless system ($\gamma=0$) with a linear quench profile ($r=1$). 
(a)~Dynamics of the stored energy $E_B(t)$ (in units of $\omega_0 F^2$) for various quench durations $\tau_Q$. 
The envelope of the oscillations grows with $\tau_Q$, contrasting with the bounded behavior of the instantaneous quench ($r=0$) shown in the inset. The dashed line in the inset corresponds to the $t^2$ scaling.
(b)~Log-log plot of the maximum stored energy $E_{B,m}$ (blue circles) and maximum power $P_{B,m}$ (red squares) as functions of the quench duration $\tau_Q$ (in units of $1/\omega_0$).  
The dotted lines represent the analytical scaling laws predicted by Eq.~(\ref{scaling}), $E_{B,m} \propto \tau_Q^{2\alpha}$ and $P_{B,m} \propto \tau_Q^{\alpha}$ with $\alpha=0.5$, showing excellent agreement with numerical simulations.
}
\end{figure}

The maximum value of the stored energy can be determined by solving $dE_B/dt=0$.  For example, for a linear quench $(r=1,\alpha=1/2)$, the first maximum is obtained numerically from Eq.~(\ref{E_B}) at $\theta_m\simeq 2.14$.  This gives
\begin{equation}
E_{B,m}^{(r=1)}\propto \tau_Q,
\qquad
P_{B,m}^{(r=1)}\propto \tau_Q^{1/2}.
\end{equation}
Figure~\ref{fig2} depicts the dynamics of the stored energy $E_B(t)$ [Fig.~\ref{fig2}(a)] and the scaling of the maximum power $P_{B,m}$ [Fig.~\ref{fig2}(b)].  In Fig.~\ref{fig2}(a), we observe that while a fast quench with small $\tau_Q$ results in coherent oscillations bounded by the maximum value in Eq.~(\ref{vanilla}), a slow quench allows the energy to grow monotonically beyond the limits of the constant-coupling case.  Figure~\ref{fig2}(b) confirms the algebraic scaling law $P_{B,m}\propto \tau_Q^\alpha$, evidenced by the linear slope in the log-log plot, where the exponent $\alpha$ approaches unity for large quench exponents $r$.

%-----------------------------------------------------------
%-----------------------------------------------------------

The reason why a slow interaction quench leads to algebraic energy scaling in a closed system can be understood from Eq.~(\ref{heis_eom}). In the decoupled case, where $g(t)=0$, the energy of the charger $E_{A}$ grows as $\sim F^2 t^2$. If the charger is immediately connected to the battery with constant coupling, $E_{B}$ loses its quadratic growth due to coherent energy oscillations between the charger and the battery. However, under a slowly quenched coupling, the battery retains quadratic growth as in Eq.~(\ref{E_B}) up to a certain time; $E_{A}$ also grows similarly. 
In this case, once the oscillatory factor in Eq.~(\ref{E_B}) starts to dominate the dynamics, $E_B$ reaches its first maximum at $t_\text{m}\sim \tau_Q^\alpha$.
This behavior results in a monotonic increase of $P_{B,m}$ with exponent $\alpha$. The fact that the energy of the decoupled charger grows quadratically, resulting in linear power growth, is exactly why $\alpha$ has an upper bound of~$1$. Hence, we conclude that continuous quench may suppress coherent energy oscillations so that coherent driving pumps the charger and the battery more effectively.

Let us now consider the extractable work stored in the battery. The ergotropy of a battery state $\rho_B$ is defined as
\begin{equation}\label{ergo}
\mathcal{W}_B
=
\mathrm{Tr}(\rho_B H_B)
-
\min_U \mathrm{Tr}(U\rho_B U^\dagger H_B),
\end{equation}
where the minimization is over unitary operations acting only on the battery. The state that minimizes the second term of Eq.~(\ref{ergo}) is referred to as the passive state. In the ideal closed bosonic model considered here, the dynamics is generated by a quadratic Hamiltonian and the battery is initially in the oscillator vacuum. The coherent drive and beam-splitter-type charger-battery interaction therefore keep the reduced battery state in a coherent state,
\begin{equation}
\rho_B(t)=|\beta(t)\rangle\langle\beta(t)|,
\qquad
E_B(t)=\omega_0|\beta(t)|^2 .
\end{equation}
Since the passive state associated with a pure coherent state of a harmonic oscillator is the vacuum, all stored energy above the ground state is extractable:
\begin{equation}
\mathcal{W}_B(t)=E_B(t),
\qquad
\frac{\mathcal{W}_B(t)}{E_B(t)}=1 .
\end{equation}
This result is consistent with previous analyses of charger-mediated quantum batteries, where correlations were shown to reduce the extractable fraction of stored energy, while coherent charging mitigates this suppression~\cite{andolina2018charger,andolina2019extractable}.

This observation also clarifies the role of the energy fluctuations. For a coherent state,
\begin{equation}
\Delta E_B^2
=
\omega_0^2|\beta(t)|^2 .
\end{equation}
Thus, larger quench exponents can increase the absolute energy fluctuations together with the peak energy and peak power. However, these fluctuations do not correspond to passive energy and do not reduce the ergotropy ratio in the ideal protocol. The appropriate operational procedure is to switch off the charger-battery coupling at the first maximum of $E_B(t)$ or $P_B(t)$, thereby preventing coherent backflow to the charger. In open-system settings, the battery state may become mixed, and incoherent or thermal excitations can reduce the ergotropy below the stored energy. For example, for a displaced thermal state,
\begin{equation}
\rho_B=D(\beta)\rho_{\rm th}D^\dagger(\beta),
\end{equation}
one can obtain
\begin{align}
E_B &=\omega_0(|\beta|^2+n_{\rm th}),\\
\mathcal{W}_B&=\omega_0|\beta|^2,\\
\frac{\mathcal{W}_B}{E_B}&=
\frac{|\beta|^2}{|\beta|^2+n_{\rm th}} .
\end{align}
Thus the ideal result $\mathcal{W}_B/E_B=1$ is robust as long as the stored excitation remains dominantly coherent.
    
The equality $\mathcal{W}_B=E_B$ holds for the ideal closed dynamics considered in this section. It relies on the fact that the stored excitation remains coherent and that the charger-battery coupling is switched off at the first charging maximum. In realistic open systems, loss and dephasing can limit the coherent amplitude and may generate mixed battery states, for which the ergotropy can be smaller than the stored energy. We therefore next examine how charger dissipation modifies the scaling window.

%-----------------------------------------------------------
\section{Effects of charger dissipation}
\label{sec:dissipation}
%-----------------------------------------------------------

In the presence of dissipation $\gamma\neq 0$, the exact solution of Eq.~(\ref{eom}) is unknown. However, analytical solution is available for $r\to\infty\,(\alpha=1)$. In this case, the coupling can be replaced by a step function $g(t)={g_{f}}\,\theta(t-\tau_Q)$. Thus, for $t\in[0,\tau_Q)$ the charger is effectively decoupled from the battery and its stored energy reads
\begin{align}\label{EA_r_infty}
    E_{A}^{(r\to\infty)}(t) &= \omega_0\frac{4F^2}{\gamma^2}\left(e^{-\gamma t/2}-1\right)^2 \nonumber\\
    &=\omega_0 F^2 t^2 + O(t^3),
\end{align}
which also grows quadratically at initial times, as in the dissipationless case.  
After the coupling is switched on at $t=\tau_Q$, the accumulated charger excitation is transferred to the battery over a coherent swap time $\delta t_{\rm sw}=O(g_f^{-1})$. Thus the first battery maximum occurs at $t_m=\tau_Q+\delta t_{\rm sw}$, not exactly at $\tau_Q$. In the scaling estimate below we focus on the regime $\tau_Q\gg g_f^{-1}$, where this additive transfer time is negligible for the $\tau_Q$-dependence of the average power. 
Consequently, the maximum average power of the battery exhibits the following scaling behavior,
\begin{equation}\label{scaling_diss}
    P_{B,m}^{(r\to\infty)}\sim \frac{1}{\tau_Q\gamma^2}\left(e^{-\gamma\tau_Q/2}-1\right)^2,
\end{equation}
while the maximum energy scales similarly with Eq.~(\ref{EA_r_infty}). 

%-----------------------------------------------------------
\begin{figure}[tb]
\centering\includegraphics[width=\columnwidth]{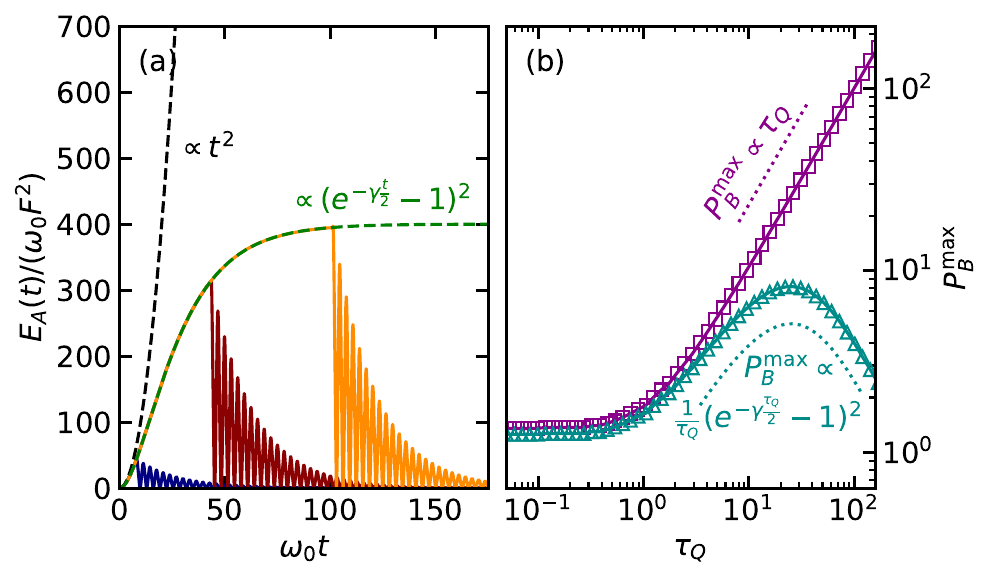}
\caption{\label{fig3} 
Effect of charger dissipation $\gamma$ on the battery performance. 
(a) Time evolution of the charger energy $E_A(t)$ (normalized by $F^2$) for some representative quench durations.  In the presence of dissipation (three solid curves), the energy saturates at long times (black dashed line), deviating from the quadratic growth observed in the lossless limit (violet dashed line). (b) Maximum battery power $P_{B,m}$ versus quench duration $\tau_Q$ (in units of $1/\omega_0$) for the lossless case (blue circles) with $\gamma = 0$ and the dissipative case (red squares) with $\gamma = 0.1$. The blue and red dotted lines are guides for the eye to emphasize the analytical scaling laws.  Note that while the algebraic scaling holds for short quenches ($\tau_Q < \tau_Q^{\text{max}}$), dissipation imposes a peak power at $\tau_Q^{\text{max}} \approx 2.513 \gamma^{-1}$, beyond which the power decays as $\tau_Q^{-1}$.
} 
\end{figure}
%-----------------------------------------------------------

It is apparent that dissipation breaks the monotonicity of $P_{B,m}$. Equation~(\ref{scaling_diss}) indicates that the power reaches the peak at 
\begin{equation}
\label{eq:tauqmax}
    \tau_Q^\text{max}
    =
    \frac{1}{\gamma}
    \left[
    -2W_{-1}\left(-\frac{1}{2\sqrt e}\right)-1
    \right]
    \approx 2.513\,\gamma^{-1}.
\end{equation}
where $W_j$ is the $j$-th branch of the Lambert $W$-function with the particular branch $j=-1$ selected as it yields the relevant real solution corresponding to the global maximum of the power function for $\tau_Q > 0$.  Thus, the peak reduces and shifts toward smaller $\tau_Q$ for stronger dissipation. For an intermediate quench duration, the linear scaling is still observed in $\pi/{g_{f}} \lessapprox \tau_Q \lessapprox \tau_Q^\text{max}$, while for a slower quench, $\tau_Q \gtrapprox \tau_Q^\text{max}$, the scaling breaks down to $P_{B,m}\propto\tau_Q^{-1}$. This is due to the fact that the charger energy $E_{A}(t)$ in Eq.~(\ref{EA_r_infty}) is bounded by a constant, $4\omega_0 F^2/\gamma^2$, for long quenches. Moreover, for quenches with a finite $r$, extensive numerical results indicate that a similar nonmonotonic scaling behavior of $P_{B,m}$ applies.
The derivation of Eqs.~(\ref{EA_r_infty})--(\ref{scaling_diss}) and the optimal quench duration in Eq.~(\ref{eq:tauqmax}) is given in Appendix~\ref{app:dissipative_step}.

In Fig.~\ref{fig3},  we show the impact of charger dissipation $\gamma$ on the charging performance. Figure~\ref{fig3}(a) displays the charger energy $E_A(t)$, which saturates at long times due to the competition between the coherent drive and dissipation.  Consequently, as shown in the right panel, the maximum battery power $P_{B,m}$ exhibits non-monotonic behavior with respect to $\tau_Q$.  There exists an optimal quench duration $\tau_Q^{\text{max}}$ that maximizes power, where beyond this point the power decays as $\tau_Q^{-1}$ as the dissipation dominates the dynamics.

%-----------------------------------------------------------
\section{Discussion}
%-----------------------------------------------------------

%-----------------------------------------------------------
\subsection{Robustness against dissipation}
%-----------------------------------------------------------

\begin{figure*}[tb]
\centering\includegraphics[clip,width=18cm]{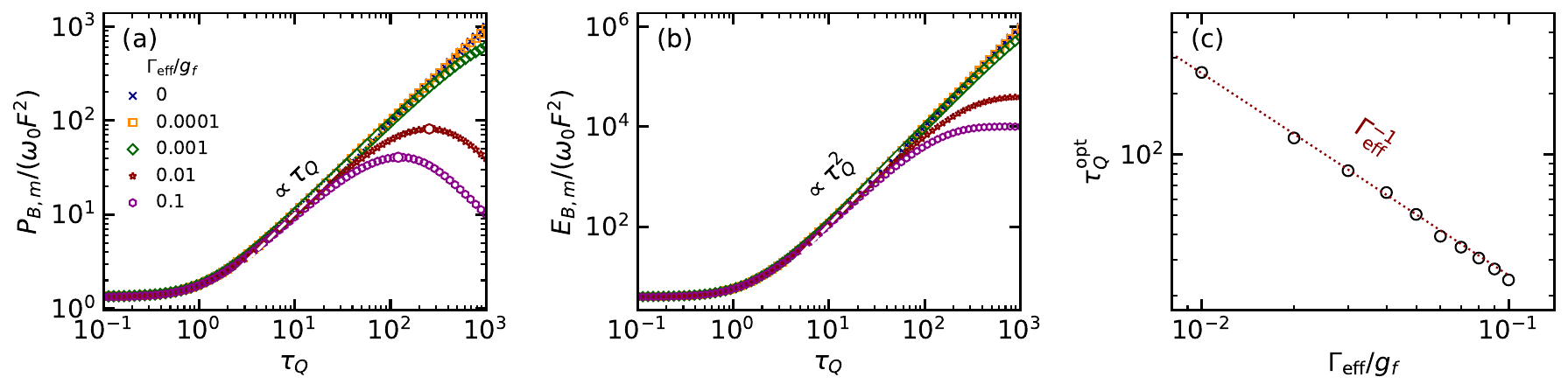}
\caption{\label{fig4} 
Robustness of the temporal scaling under effective charger decoherence. (a) Maximum battery power $P_{B,m}$ as a function of the quench duration $\tau_Q$ for several values of the effective charger loss/decoherence rate $\Gamma_{\rm eff}$. The dashed line indicates the closed-system algebraic trend $P_{B,m}\propto \tau_Q^{r/(r+1)}$, evaluated here for the quasi-step ramp $r=1000$. Weak dissipation preserves an algebraic window, while stronger dissipation causes an earlier turnover at large $\tau_Q$. (b) Corresponding maximum stored battery energy $E_{B,m}$. The dashed guide indicates the lossless scaling trend. The departure from this trend reflects the saturation of the charger amplitude under loss. (c) Optimal quench duration $\tau_Q^{\rm opt}$, extracted from the maximum of $P_{B,m}(\tau_Q)$, as a function of $\Gamma_{\rm eff}/g_f$ within $[10^{-2}, 10^{-1}]$.  The dashed line shows a guide proportional to $\Gamma_{\rm eff}^{-1}$.  The results indicate that dissipation converts the closed-system algebraic law into a finite-time scaling window followed by a dissipation-controlled optimum.}
\end{figure*}

It should be noted that the algebraic enhancement can be modified, but not immediately destroyed, by realistic loss and decoherence.  To quantify this effect, we performed an additional open-system benchmark by replacing the charger loss rate $\gamma$ in Eq.~(\ref{lindblad}) with an effective amplitude-decay rate $\Gamma_{\rm eff}$.
Equivalently, $\Gamma_{\rm eff}$ enters the first-moment equations as
\begin{align}
\dot A(t)&=-i[g(t)B(t)+F]-\frac{\Gamma_{\rm eff}}{2}A(t),\\
\dot B(t)&=-ig(t)A(t),
\end{align}
so that it phenomenologically represents all processes that reduce the coherent charger amplitude.  This phenomenological parameter represents the dominant mechanism that limits the coherent amplitude accumulated in the charger mode, such as photon loss or amplitude damping in superconducting resonators. 

The resulting maximum battery power and stored energy are shown in Fig.~\ref{fig4}. In Fig.~\ref{fig4}(a), $P_{B,m}$ initially follows the same algebraic trend as in the lossless case. This demonstrates that the enhancement is not a purely closed-system artifact. 
However, the algebraic growth with a positive exponent holds only within a finite window of ramp durations. For sufficiently large $\tau_Q$, the charger amplitude becomes loss-limited, and further increasing the quench duration no longer improves the charging power. This produces a dissipative turnover in $P_{B,m}$, whose position shifts to smaller $\tau_Q$ as $\Gamma_{\rm eff}$ is increased.

The same interpretation is supported by the stored-energy curves in Fig.~\ref{fig4}(b). For weak dissipation, $E_{B,m}$ follows the lossless scaling over a broad interval of $\tau_Q$, while stronger dissipation causes an earlier departure from the algebraic trend and eventually a saturation of the stored energy. The saturation reflects the balance between the drive-induced growth of the charger amplitude and the dissipative loss of coherent excitation. Thus, dissipation does not change the microscopic origin of the enhancement, but it bounds the time over which the enhancement can be accumulated. 
Figure~\ref{fig4}(c) makes this bound explicit by extracting the optimal quench duration $\tau_Q^{\rm opt}$ from the maximum of $P_{B,m}(\tau_Q)$.  We include several values of $\Gamma_{\rm eff}/g_f$ in the range $10^{-2}\leq \Gamma_{\rm eff}/g_f\leq 10^{-1}$.  This interval is chosen because it corresponds to the range most relevant to the order-of-magnitude experimental estimate and because the dissipative turnover is clearly visible within the numerical time window.
For much smaller effective loss, such as $\Gamma_{\rm eff}/g_f\lesssim 10^{-3}$, the turnover is shifted to much longer ramp durations and the dynamics is nearly indistinguishable from the lossless algebraic scaling over the plotted interval.
The numerical data show the approximate relation
\begin{equation}
\tau_Q^{\rm opt}\propto \Gamma_{\rm eff}^{-1},
\label{tau_opt}
\end{equation}
which is the expected scaling, consistent with Eq.~(\ref{eq:tauqmax}), if the algebraic window closes when the ramp duration becomes comparable to the effective decoherence time.
Therefore, the experimentally relevant prediction is not an indefinitely valid algebraic law, but a finite algebraic regime followed by a loss-induced optimum.

The robustness criterion is platform dependent. In superconducting bosonic circuits, $\Gamma_{\rm eff}$ can be interpreted primarily as microwave photon loss and dephasing, so the observation of the algebraic regime requires $\tau_Q$ to remain below the relevant resonator coherence time. In organic microcavities, the role of decoherence can be more subtle. Experiments on organic microcavity quantum batteries have shown that the stored energy can be monitored through transient reflectivity, with the signal proportional to the excited molecular population, and that charging dynamics, stored energy, and charging power can be extracted from ultrafast pump-probe measurements~\cite{quach2022superabsorption,Hymas2026SuperextensiveElectricalPower}. In such systems, dephasing may also assist energy retention by transferring population from optically bright states to dark states, thereby suppressing rapid re-emission~\cite{quach2022superabsorption}. Hence, realistic open-system effects restrict and reshape the observable scaling window, but they do not necessarily preclude the observation of algebraic or collective charging signatures.

As an order-of-magnitude benchmark, we may use the cavity lifetime $T=120\text{--}306\,\mathrm{fs}$ reported in the organic microcavity experiment of Quach et al.~\cite{quach2022superabsorption}. This gives a cavity decay scale $T^{-1}=3.27\text{--}8.33\,\mathrm{ps}^{-1}$. Identifying this scale with $\Gamma_{\rm eff}$, and using the corresponding light-matter coupling scale only as a platform-dependent estimate, places $\Gamma_{\rm eff}/g_f$ between the representative curves $\Gamma_{\rm eff}/g_f=0.01$ and $0.1$ in Fig.~\ref{fig4}(a) and~\ref{fig4}(b). It should be noted that the comparison is a kind of dimensional benchmark.  We do not intend it as a direct fit to the molecular microcavity experiment since that platform realizes a collective Dicke-type battery rather than the ramped oscillator-oscillator protocol studied here. Within this range, Fig.~\ref{fig4} suggests that sizeable enhancements of the maximum stored energy and charging power can remain observable before the dissipative turnover occurs.

%-----------------------------------------------------------
\subsection{Mapping to a driven Tavis-Cummings battery}
%-----------------------------------------------------------

An important aspect of our quantum battery model is that it can be related to a finite collective-spin battery in the low-excitation regime. This connection follows from the Holstein-Primakoff representation of the collective battery spin,
\begin{equation}
S^{+}=b^{\dagger}\sqrt{2s-b^\dagger b},\qquad
S^{-}=\sqrt{2s-b^\dagger b}\,b ,
\end{equation}
where $s$ is the collective spin quantum number and the corresponding symmetric spin Hilbert space has dimension $2s+1$. In the low-excitation limit $ \langle b^\dagger b\rangle\ll 2s$, one may approximate
\begin{equation}
S^{+}\simeq \sqrt{2s}\,b^\dagger,\qquad
S^{-}\simeq \sqrt{2s}\,b .
\end{equation}
Under this approximation, the Hamiltonian in Eq.~(\ref{model}) can be written as
\begin{equation}
    H(t) \simeq \frac{g(t)}{\sqrt{2s}}\left(a S^+ + a^\dagger S^-\right)
    + F(a+a^\dagger),
\end{equation}
which has the form of a coherently driven Tavis-Cummings quantum battery~\cite{lu2021optimal,quach2022superabsorption,yang2024optimal,canzio2025single,Hymas2026SuperextensiveElectricalPower}, or equivalently the rotating-wave approximation of the driven Dicke model~\cite{dicke1954coherence}.  Thus, molecular microcavity quantum batteries, where a cavity mode couples collectively to a large ensemble of molecular excitations, provide a related experimental platform~\cite{quach2022superabsorption,Hymas2026SuperextensiveElectricalPower}.
This mapping suggests that the temporal algebraic enhancement induced by slow quenching can also appear in collective-spin quantum batteries, provided the dynamics remains within the low-excitation regime where the Holstein-Primakoff approximation is valid.

The ramped coupling $g(t)$ is also reminiscent of protocols in which a driven collective light-matter system is swept across a superradiant critical region~\cite{bastidas2012nonequilibrium,jara2024apparent}. In such finite-rate critical sweeps, nonadiabatic excitations may be discussed in the language of the Kibble-Zurek mechanism~\cite{kibble1976topology,kibble1980some,zurek1985cosmological,zurek1996cosmological,campo2014universality,dziarmaga2010dynamics}. However, the temporal scaling found in the present bosonic battery should be distinguished from equilibrium finite-size scaling near a superradiant critical point. The algebraic dependence on the quench duration $\tau_Q$ does not rely on tuning the instantaneous coupling $g(t)$ through a specific critical value $g_c$. Rather, it follows from the coherent build-up of the charger amplitude during the ramp and the subsequent beam-splitter-mediated transfer to the battery. Hence the relevant control parameter is the ramp duration, not the distance from a superradiant critical coupling. It should be noted that in the rotating frame considered, superradiance transition occurs at $g_c=\sqrt{\Delta_a \Delta_b}$ where $\Delta_{a,b}=\omega_{a,b}-\omega_d$ is the charger (battery)-drive detuning. 
In the resonant rotating-frame limit considered here, this critical scale collapses to zero, so the protocol does not involve crossing a finite nonzero critical coupling. The observed exponent is therefore controlled by the ramp time and the coherent build-up of the charger amplitude, not by the distance from a finite superradiant threshold. For detuned systems, or when counter-rotating terms and dissipation are included, the relation to superradiant criticality can change and requires a separate finite-size and open-system analysis~\cite{kirton2019introduction,keeling2010collective,larson2017some,soriente2018dissipation}.

The connection to the Tavis-Cummings or Dicke-type description should therefore be understood as a low-excitation mapping between a collective spin ensemble and an effective bosonic mode. In this regime, the bosonic battery energy corresponds to collective spin excitation energy. The approximation remains valid when the maximum number of stored battery excitations satisfies
\begin{equation}
n_{B,m}\ll 2s .
\end{equation}
Equivalently, for a fully symmetric ensemble of $N=2s$ two-level systems, the condition is $n_{B,m}\ll N$. In this regime, the oscillator model captures the leading charging dynamics, and the temporal scaling is essentially independent of crossing a specific superradiant critical point. When $n_{B,m}$ becomes comparable to $2s$, spin saturation and finite-size corrections become important. These corrections can alter the scaling and may obscure any relation to the superradiant phase transition. A systematic finite-size scaling analysis near the critical region is therefore beyond the scope of the present work.

%-----------------------------------------------------------
\subsection{Possible experimental realizations and signatures}
%-----------------------------------------------------------

The bosonic version of our model can be implemented in superconducting circuit quantum electrodynamics (circuit-QED) architectures, where the charger and battery may be encoded in two microwave resonators or high-$Q$ three-dimensional cavities. The essential experimental requirement is not only a tunable coupling, but a calibrated waveform-controlled beam-splitter interaction,
\begin{equation}
H_{\rm int}(t)=g(t)(a^\dagger b+ab^\dagger),
\end{equation}
with the prescribed ramp profile $g(t)=g_{\max}(t/\tau_Q)^r$. Such control is within the capabilities of present circuit-QED devices. 
For example, superconducting quantum interference device (SQUID)-based and three-wave-mixing couplers have been used to generate programmable beam-splitter interactions between superconducting bosonic modes, with fast coherent exchange, large on-off ratios, and weak additional decoherence channels~\cite{Lu2023ParametricBeamsplitting,Chapman2023HighOnOffBeamsplitter}. Related experiments have also demonstrated flux-tunable coupling between superconducting transmission-line resonators mediated by a radio-frequency SQUID (rf-SQUID)~\cite{Wulschner2016TunableCoupling}.
In these platforms, the ramp $g(t)$ can be implemented by shaping the amplitude of the parametric pump or the external flux waveform after calibrating the beam-splitter rate against the applied control amplitude.

The most direct experimental signature of the temporal extensivity predicted here is a kind of scaling measurement similar to those of Figs.~\ref{fig3}--\ref{fig4}.  One would repeat the charging protocol for a sequence of quench durations $\tau_Q$, reconstruct the battery occupation $n_B(t)=\langle b^\dagger b\rangle$, and extract the first maximum $E_{B,m}=\omega_0 n_{B,m}$ together with the corresponding maximum average power $P_{B,m}=E_{B,m}/t_m$. In the weak-loss regime, a log-log plot of $P_{B,m}$ versus $\tau_Q$ should display the algebraic scaling $P_{B,m}\propto \tau_Q^\alpha$, with $\alpha$ determined by the ramp exponent $r$. The same experiment can also test the dissipative prediction by engineering the charger linewidth $\gamma$, or more generally the effective amplitude-decay rate $\Gamma_{\rm eff}$. Increasing $\gamma$ should convert the monotonic algebraic growth into a nonmonotonic dependence on $\tau_Q$, producing a finite optimal quench duration $\tau_{Q}^{\rm opt}$, as in Fig.~\ref{fig3}(b). The associated time-domain data should show the charger occupation saturating at long times, as in Fig.~\ref{fig3}(a), rather than continuing the quadratic growth expected in the lossless limit.

The mapping to the Tavis–Cummings model allows our protocol to be realized in microcavity quantum batteries, where a cavity mode is collectively coupled to a large ensemble of molecular excitations forming an effective collective spin. Such platforms have recently been demonstrated experimentally in Refs. \citenum{quach2022superabsorption} and \citenum{Hymas2026SuperextensiveElectricalPower}. Our model requires additional coherent drive $F$, which corresponds to coherent optical pumping of the cavity mode, and a controllable light-matter coupling $g(t)$. 

The relevant observables are accessible with standard microwave-cavity tools, including calibrated heterodyne detection of the output field, cavity-state tomography, and photon-number-resolved measurements using dispersively coupled ancilla qubits. Organic microcavity quantum batteries and NMR quantum-battery experiments provide complementary evidence that charging dynamics, stored energy, and power scaling can already be measured experimentally~\cite{quach2022superabsorption,Joshi2022NMRQuantumBattery}. Nevertheless, superconducting bosonic circuits appear to be the most direct platform for implementing the specific ramped oscillator-oscillator protocol proposed here.

%-----------------------------------------------------------
\section{Conclusions}
%-----------------------------------------------------------
We have demonstrated that a slow-quench charging protocol in a coherently driven oscillator battery leads to algebraic scaling of both the maximum stored energy and charging power with the quench duration, $E_{B,m}\propto \tau_Q^{2\alpha}$ and $P_{B,m} \propto \tau_Q^{\alpha}$. This temporal extensivity allows the battery to overcome the energy limitations imposed by coherent Rabi oscillations in standard constant-coupling schemes. Our results suggest an optimal charging protocol in which the interaction is turned off at $t_m$ to store the maximum energy, which is fully extractable as ergotropy in the ideal closed model. Although higher quench exponents $r$ yield better scaling and higher power, they induce larger energy fluctuations and require precise timing near $\tau_Q$ due to the sharp peak. Furthermore, the inclusion of dissipation reveals a finite optimal quench duration, providing a practical bound for experimental implementations. Finally, the mapping to the Tavis-Cummings model suggests that this algebraic scaling may extend to collective-spin quantum batteries in the low-excitation regime.

%-----------------------------------------------------------
\section*{Data and Code Availability}
%-----------------------------------------------------------

Additional data, notes, and Python codes that can be used to reproduce our results are available to download from \url{https://github.com/BRIN-Q/slow-quench-qbattery}

%-----------------------------------------------------------
\begin{acknowledgments}
D.D. was supported by the APCTP Young Scientist Training (YST) program through the Science and Technology Promotion Fund and the Lottery Fund of the Korean Government and the Korean Local governments (Gyeongsangbuk-do Province and Pohang city).  He also acknowledges the BRIN Postdoctoral Program in 2023, during which the ideas for this work emerged.  All authors acknowledge QuasiLab and Mahameru BRIN for their mini-cluster and HPC facilities.
\end{acknowledgments}
%-----------------------------------------------------------
\appendix
%-----------------------------------------------------------

%-----------------------------------------------------------
\section{Constant-coupling solution}
\label{app:constant_coupling}

In this appendix, we derive Eq.~(\ref{vanilla}) for the constant-coupling protocol. We set $r=0$, $g(t)=g_f$, and $\gamma=0$. The Heisenberg equations for the first moments are
\begin{align}
\dot a(t)&=-i g_f b(t)-iF,\\
\dot b(t)&=-i g_f a(t).
\end{align}
For the initial vacuum state, the dynamics generated by the quadratic Hamiltonian and the coherent drive keeps the state Gaussian and coherent. Therefore, the stored energy can be obtained from the coherent amplitudes [Eq.~\eqref{eq:coh-amplitudes}],
\begin{displaymath}
A(t)=\langle a(t)\rangle,\qquad B(t)=\langle b(t)\rangle .
\end{displaymath}
The corresponding equations are
\begin{align}
\dot A(t)&=-i g_f B(t)-iF,\\
\dot B(t)&=-i g_f A(t),
\end{align}
with initial conditions $A(0)=B(0)=0$.

Differentiating the second equation gives
\begin{equation}
\ddot B(t)=-i g_f \dot A(t)
=-g_f^2 B(t)-g_f F .
\end{equation}
Thus
\begin{equation}
\ddot B(t)+g_f^2 B(t)=-g_f F .
\end{equation}
Solving with $B(0)=0$ and $\dot B(0)=0$ yields
\begin{equation}
B(t)=\frac{F}{g_f}\left[\cos(g_f t)-1\right]
=-\frac{2F}{g_f}\sin^2\left(\frac{g_f t}{2}\right).
\end{equation}
Since the battery is in a coherent state, its occupation is
\begin{equation}
n_B(t)=\langle b^\dagger b\rangle=|B(t)|^2 .
\end{equation}
The battery energy is therefore
\begin{equation}
E_B^{(r=0)}(t)
=
\omega_0 |B(t)|^2
=
4\frac{\omega_0 F^2}{g_f^2}
\sin^4\left(\frac{g_f t}{2}\right),
\end{equation}
which is Eq.~(\ref{vanilla}). The first maximum occurs at $g_f t=\pi$, giving
\begin{align}
t_m^{(r=0)}&=\frac{\pi}{g_f},\\
E_{B,m}^{(r=0)}&=4\frac{\omega_0 F^2}{g_f^2},\\
P_{B,m}^{(r=0)}&=\frac{4\omega_0F^2}{\pi g_f}.
\end{align}
%-----------------------------------------------------------

\section{Slow-quench solution and algebraic scaling}
\label{app:slow_quench_derivation}

Here we give the analytical steps leading to Eq.~(\ref{E_B}) and the scaling relation in Eq.~(\ref{scaling}). We consider the closed system, $\gamma=0$, during the ramp interval $0\leq t\leq \tau_Q$. The coupling is
\begin{equation}
g(t)=k t^r,
\qquad
k=\frac{g_f}{\tau_Q^r}.
\end{equation}
For the coherent amplitudes
\begin{displaymath}
A(t)=\langle a(t)\rangle,\qquad B(t)=\langle b(t)\rangle ,
\end{displaymath}
the Heisenberg equations become
\begin{align}
\dot A(t)&=-i k t^r B(t)-iF,\\
\dot B(t)&=-i k t^r A(t).
\end{align}
Eliminating $A(t)$ gives
\begin{equation}
t\ddot B(t)-r\dot B(t)+k^2 t^{2r+1}B(t)=-kF t^{r+1},
\end{equation}
which is Eq.~(\ref{eom}) for $\gamma=0$.

The same result can be obtained more directly by introducing the instantaneous normal-mode amplitudes
\begin{equation}
C_\pm(t)=\frac{A(t)\pm B(t)}{\sqrt{2}} .
\end{equation}
Adding and subtracting the equations for $A(t)$ and $B(t)$, one obtains
\begin{equation}
\dot C_\pm(t)=\mp i g(t)C_\pm(t)-\frac{iF}{\sqrt{2}},
\end{equation}
with initial conditions $C_\pm(0)=0$.  The solution is
\begin{equation}
C_\pm(t)
=
-\frac{iF}{\sqrt{2}}
e^{\mp i\Phi(t)}
\int_0^t e^{\pm i\Phi(t')}dt',
\end{equation}
where
\begin{equation}
\Phi(t)=\int_0^t g(t')dt'
=
\frac{k t^{r+1}}{r+1}.
\end{equation}
The battery amplitude is
\begin{equation}
B(t)=\frac{C_+(t)-C_-(t)}{\sqrt{2}} .
\end{equation}
Combining the two normal-mode contributions gives
\begin{equation}
B(t)
=
-F\int_0^t
\sin\left[\Phi(t)-\Phi(t')\right]dt' .
\end{equation}
Substituting the ramp profile into $\Phi(t)$, we obtain
\begin{equation}
B(t)
=
-F\int_0^t
\sin\left[
\frac{k}{1+r}
\left(t^{1+r}-{t'}^{1+r}\right)
\right]dt' .
\end{equation}
Setting $t'=ut$ gives
\begin{equation}
B(t)
=
-Ft\int_0^1
\sin\left[
\frac{k t^{1+r}}{1+r}
\left(1-u^{1+r}\right)
\right]du .
\end{equation}
With
\begin{equation}
\theta(t)=k t^{1+r}
=
\frac{g_f t^{1+r}}{\tau_Q^r},
\end{equation}
this becomes
\begin{equation}
B(t)
=
-Ft\int_0^1
\sin\left[
\frac{\theta(t)}{1+r}
\left(1-u^{1+r}\right)
\right]du .
\end{equation}
Since the battery remains in a coherent state, $E_B(t)=\omega_0|B(t)|^2$. Therefore,
\begin{equation}
\label{scalingderivs}
E_B(t)
=
\omega_0F^2t^2
\left[
\int_0^1
\sin\left(
\frac{\theta(t)}{1+r}
\left[1-u^{1+r}\right]
\right)du
\right]^2 ,
\end{equation}
which is Eq.~(\ref{E_B}). 
Since Eq.~\eqref{scalingderivs} is the solution during the ramp interval, the above evaluation is valid when the first maximum occurs before the end of the ramp, $t_m\leq\tau_Q$. Equivalently, $\theta_m\leq\theta(\tau_Q)=g_f\tau_Q$, which gives the slow-quench condition $\tau_Q\geq\theta_m/g_f$. When this condition is not satisfied, the first maximum occurs after the ramp has ended and the scaling derived below no longer applies.

Under the abovementioned condition, the expression in Eq.~\eqref{scalingderivs} exposes the scaling with $\tau_Q$.  The integral is a dimensionless function of $\theta(t)$ and $r$ only.  Thus one can write
\begin{equation}
B(t)=Ft\,\mathcal{F}_r[\theta(t)],
\end{equation}
where
\begin{equation}
\mathcal{F}_r[\theta]
=
-\int_0^1
\sin\left[
\frac{\theta}{1+r}
\left(1-u^{1+r}\right)
\right]du .
\end{equation}
The first maximum of the battery energy occurs at
\begin{equation}
\theta(t_m)=\theta_m ,
\end{equation}
where $\theta_m$ is determined by $r$, but is independent of $\tau_Q$.  Since
\begin{equation}
\theta_m=k t_m^{1+r}
=
\frac{g_f t_m^{1+r}}{\tau_Q^r},
\end{equation}
we obtain
\begin{equation}
t_m
=
\left(\frac{\theta_m}{g_f}\right)^{1/(1+r)}
\tau_Q^{r/(1+r)} .
\end{equation}
Defining
\begin{equation}
\alpha=\frac{r}{r+1},
\end{equation}
we have
\begin{equation}
t_m\propto \tau_Q^\alpha .
\end{equation}
At the first maximum,
\begin{equation}
E_{B,m}
=
\omega_0F^2 t_m^2
\left|\mathcal{F}_r(\theta_m)\right|^2
\propto \tau_Q^{2\alpha}.
\end{equation}
The corresponding maximum average power is
\begin{equation}
P_{B,m}
=
\frac{E_{B,m}}{t_m}
\propto
\tau_Q^\alpha .
\end{equation}
This derivation thus proves Eq.~(\ref{scaling}). Since $r\in[0,\infty)$, the exponent satisfies
\begin{equation}
0\leq \alpha\leq 1 .
\end{equation}

%-----------------------------------------------------------
\section{Dissipative step-quench limit}
\label{app:dissipative_step}
%-----------------------------------------------------------

In this appendix, we derive the analytical dissipative result used in Sec.~\ref{sec:dissipation}. We focus on the limit $r\to\infty$, for which the ramped coupling becomes a delayed step function,
\begin{equation}
g(t)=g_f\,\Theta(t-\tau_Q).
\end{equation}
For $0\leq t<\tau_Q$, the charger and battery are effectively decoupled. The battery remains in the vacuum state, while the charger is driven and damped according to
\begin{equation}
\dot A(t)=-iF-\frac{\gamma}{2}A(t),
\end{equation}
where $A(t)=\langle a(t)\rangle$. The initial condition is $A(0)=0$. Solving this first-order equation gives
\begin{equation}
A(t)
=
-\frac{2iF}{\gamma}\left(1-e^{-\gamma t/2}\right).
\end{equation}
Since the charger remains in a coherent state, its energy is
\begin{equation}
E_A(t)=\omega_0 |A(t)|^2
=
\omega_0\frac{4F^2}{\gamma^2}
\left(1-e^{-\gamma t/2}\right)^2 .
\end{equation}
Equivalently,
\begin{equation}
E_A(t)
=
\omega_0\frac{4F^2}{\gamma^2}
\left(e^{-\gamma t/2}-1\right)^2 ,
\end{equation}
which is Eq.~(\ref{EA_r_infty}). For short times, expanding the exponential gives
\begin{equation}
1-e^{-\gamma t/2}
=
\frac{\gamma t}{2}+O(t^2),
\end{equation}
and therefore
\begin{equation}
E_A(t)=\omega_0F^2t^2+O(t^3).
\end{equation}
Thus, the initial quadratic energy growth is the same as in the lossless case.

At $t=\tau_Q$, the coupling is switched on. In the ideal step-quench limit, the battery can receive the charger energy accumulated during the decoupled interval. Therefore, up to a prefactor set by the subsequent coherent swap, the maximum battery energy scales as
\begin{equation}
E_{B,m}^{(r\to\infty)}
\propto
\omega_0\frac{4F^2}{\gamma^2}
\left(1-e^{-\gamma \tau_Q/2}\right)^2 .
\end{equation}
The corresponding maximum average power is obtained by dividing by the charging time. More precisely, after the switch-on at $t=\tau_Q$, the coherent transfer from the charger to the battery requires an additional swap time $\delta t_{\rm sw}=O(g_f^{-1})$, so that $t_m=\tau_Q+\delta t_{\rm sw}$. Since this appendix is concerned with the scaling with $\tau_Q$, we work in the regime $\tau_Q\gg g_f^{-1}$ and neglect this additive transfer time, writing $t_m\simeq \tau_Q$. Thus,
\begin{equation}
P_{B,m}^{(r\to\infty)}
\propto
\frac{1}{\tau_Q\gamma^2}
\left(1-e^{-\gamma \tau_Q/2}\right)^2 .
\end{equation}
This gives Eq.~(\ref{scaling_diss}), apart from an overall constant independent of $\tau_Q$.

We now determine the value of $\tau_Q$ that maximizes this expression. Since the prefactor $1/\gamma^2$ is independent of $\tau_Q$, we maximize
\begin{equation}
f(\tau_Q)=
\frac{\left(1-e^{-\gamma \tau_Q/2}\right)^2}{\tau_Q}.
\end{equation}
It is useful to define the dimensionless variable
\begin{equation}
x=\frac{\gamma\tau_Q}{2}.
\end{equation}
Then
\begin{equation}
f(\tau_Q)
=
\frac{\gamma}{2}
\frac{(1-e^{-x})^2}{x}.
\end{equation}
Thus, the maximization reduces to maximizing
\begin{equation}
h(x)=\frac{(1-e^{-x})^2}{x}.
\end{equation}
The stationarity condition $dh/dx=0$ gives
\begin{equation}
\frac{2(1-e^{-x})e^{-x}}{x}
-
\frac{(1-e^{-x})^2}{x^2}
=0 .
\end{equation}
For $x>0$, we can divide by $1-e^{-x}$, obtaining
\begin{equation}
2xe^{-x}=1-e^{-x}.
\end{equation}
Equivalently,
\begin{equation}
e^x=2x+1.
\end{equation}
To solve this equation in terms of the Lambert $W$-function, define
\begin{equation}
y=x+\frac{1}{2}.
\end{equation}
Then $2x+1=2y$, and the equation becomes
\begin{equation}
e^{y-1/2}=2y .
\end{equation}
Rearranging,
\begin{equation}
y e^{-y}=\frac{1}{2\sqrt e}.
\end{equation}
Multiplying by $-1$, we obtain
\begin{equation}
(-y)e^{-y}=-\frac{1}{2\sqrt e}.
\end{equation}
Therefore,
\begin{equation}
-y
=
W_j\left(-\frac{1}{2\sqrt e}\right),
\end{equation}
or
\begin{equation}
x
=
-y+\frac{1}{2}
=
-W_j\left(-\frac{1}{2\sqrt e}\right)-\frac{1}{2}.
\end{equation}
The physically relevant nonzero maximum corresponds to the $j=-1$ branch since the principal branch gives the trivial solution $x=0$. Hence
\begin{equation}
x_{\rm max}
=
-W_{-1}\left(-\frac{1}{2\sqrt e}\right)-\frac{1}{2}.
\end{equation}
Returning to $\tau_Q=2x/\gamma$, we obtain
\begin{equation}
\tau_Q^{\rm max}
=
\frac{1}{\gamma}
\left[
-2W_{-1}\left(-\frac{1}{2\sqrt e}\right)-1
\right]
\simeq
2.513\,\gamma^{-1}.
\end{equation}
This result is equivalent to Eq.~\eqref{eq:tauqmax}. For $\tau_Q\ll \gamma^{-1}$, the expansion $1-e^{-\gamma\tau_Q/2}\simeq \gamma\tau_Q/2$ gives
\begin{equation}
P_{B,m}^{(r\to\infty)}
\propto
\tau_Q ,
\end{equation}
which recovers the lossless algebraic scaling with $\alpha=1$. For $\tau_Q\gg\gamma^{-1}$, the exponential term saturates, and
\begin{equation}
P_{B,m}^{(r\to\infty)}
\propto
\tau_Q^{-1}.
\end{equation}
Thus, charger dissipation converts the monotonic algebraic enhancement into a finite scaling window followed by a dissipative decay.

%-------------------------------------------------
%\bibliography{refs-new} 

\begin{thebibliography}{78}%
\makeatletter
\providecommand \@ifxundefined [1]{%
 \@ifx{#1\undefined}
}%
\providecommand \@ifnum [1]{%
 \ifnum #1\expandafter \@firstoftwo
 \else \expandafter \@secondoftwo
 \fi
}%
\providecommand \@ifx [1]{%
 \ifx #1\expandafter \@firstoftwo
 \else \expandafter \@secondoftwo
 \fi
}%
\providecommand \natexlab [1]{#1}%
\providecommand \enquote  [1]{``#1''}%
\providecommand \bibnamefont  [1]{#1}%
\providecommand \bibfnamefont [1]{#1}%
\providecommand \citenamefont [1]{#1}%
\providecommand \href@noop [0]{\@secondoftwo}%
\providecommand \href [0]{\begingroup \@sanitize@url \@href}%
\providecommand \@href[1]{\@@startlink{#1}\@@href}%
\providecommand \@@href[1]{\endgroup#1\@@endlink}%
\providecommand \@sanitize@url [0]{\catcode `\\12\catcode `\$12\catcode
  `\&12\catcode `\#12\catcode `\^12\catcode `\_12\catcode `\%12\relax}%
\providecommand \@@startlink[1]{}%
\providecommand \@@endlink[0]{}%
\providecommand \url  [0]{\begingroup\@sanitize@url \@url }%
\providecommand \@url [1]{\endgroup\@href {#1}{\urlprefix }}%
\providecommand \urlprefix  [0]{URL }%
\providecommand \Eprint [0]{\href }%
\providecommand \doibase [0]{https://doi.org/}%
\providecommand \selectlanguage [0]{\@gobble}%
\providecommand \bibinfo  [0]{\@secondoftwo}%
\providecommand \bibfield  [0]{\@secondoftwo}%
\providecommand \translation [1]{[#1]}%
\providecommand \BibitemOpen [0]{}%
\providecommand \bibitemStop [0]{}%
\providecommand \bibitemNoStop [0]{.\EOS\space}%
\providecommand \EOS [0]{\spacefactor3000\relax}%
\providecommand \BibitemShut  [1]{\csname bibitem#1\endcsname}%
\let\auto@bib@innerbib\@empty
%</preamble>
\bibitem [{\citenamefont {Gemmer}\ \emph {et~al.}(2009)\citenamefont {Gemmer},
  \citenamefont {Michel},\ and\ \citenamefont {Mahler}}]{gemmer2009quantum}%
  \BibitemOpen
  \bibfield  {author} {\bibinfo {author} {\bibfnamefont {J.}~\bibnamefont
  {Gemmer}}, \bibinfo {author} {\bibfnamefont {M.}~\bibnamefont {Michel}},\
  and\ \bibinfo {author} {\bibfnamefont {G.}~\bibnamefont {Mahler}},\ }\href
  {https://doi.org/10.1007/978-3-540-70510-9} {\emph {\bibinfo {title} {Quantum
  thermodynamics: \text{Emergence} of thermodynamic behavior within composite
  quantum systems}}},\ Vol.\ \bibinfo {volume} {784}\ (\bibinfo  {publisher}
  {Springer},\ \bibinfo {year} {2009})\BibitemShut {NoStop}%
\bibitem [{\citenamefont {Kosloff}(2013)}]{kosloff2013quantum}%
  \BibitemOpen
  \bibfield  {author} {\bibinfo {author} {\bibfnamefont {R.}~\bibnamefont
  {Kosloff}},\ }\bibfield  {title} {\bibinfo {title} {Quantum thermodynamics:
  \text{A} dynamical viewpoint},\ }\href {https://doi.org/10.3390/e15062100}
  {\bibfield  {journal} {\bibinfo  {journal} {Entropy}\ }\textbf {\bibinfo
  {volume} {15}},\ \bibinfo {pages} {2100} (\bibinfo {year}
  {2013})}\BibitemShut {NoStop}%
\bibitem [{\citenamefont {Millen}\ and\ \citenamefont
  {Xuereb}(2016)}]{millen2016perspective}%
  \BibitemOpen
  \bibfield  {author} {\bibinfo {author} {\bibfnamefont {J.}~\bibnamefont
  {Millen}}\ and\ \bibinfo {author} {\bibfnamefont {A.}~\bibnamefont
  {Xuereb}},\ }\bibfield  {title} {\bibinfo {title} {Perspective on quantum
  thermodynamics},\ }\href {https://doi.org/10.1088/1367-2630/18/1/011002}
  {\bibfield  {journal} {\bibinfo  {journal} {New J. Phys.}\ }\textbf {\bibinfo
  {volume} {18}},\ \bibinfo {pages} {011002} (\bibinfo {year}
  {2016})}\BibitemShut {NoStop}%
\bibitem [{\citenamefont {Vinjanampathy}\ and\ \citenamefont
  {Anders}(2016)}]{vinjanampathy2016quantum}%
  \BibitemOpen
  \bibfield  {author} {\bibinfo {author} {\bibfnamefont {S.}~\bibnamefont
  {Vinjanampathy}}\ and\ \bibinfo {author} {\bibfnamefont {J.}~\bibnamefont
  {Anders}},\ }\bibfield  {title} {\bibinfo {title} {Quantum thermodynamics},\
  }\href {https://doi.org/10.1080/00107514.2016.1201896} {\bibfield  {journal}
  {\bibinfo  {journal} {Contemp. Phys.}\ }\textbf {\bibinfo {volume} {57}},\
  \bibinfo {pages} {545} (\bibinfo {year} {2016})}\BibitemShut {NoStop}%
\bibitem [{\citenamefont {Goold}\ \emph {et~al.}(2016)\citenamefont {Goold},
  \citenamefont {Huber}, \citenamefont {Riera}, \citenamefont {Del~Rio},\ and\
  \citenamefont {Skrzypczyk}}]{goold2016role}%
  \BibitemOpen
  \bibfield  {author} {\bibinfo {author} {\bibfnamefont {J.}~\bibnamefont
  {Goold}}, \bibinfo {author} {\bibfnamefont {M.}~\bibnamefont {Huber}},
  \bibinfo {author} {\bibfnamefont {A.}~\bibnamefont {Riera}}, \bibinfo
  {author} {\bibfnamefont {L.}~\bibnamefont {Del~Rio}},\ and\ \bibinfo {author}
  {\bibfnamefont {P.}~\bibnamefont {Skrzypczyk}},\ }\bibfield  {title}
  {\bibinfo {title} {The role of quantum information in
  thermodynamics---\text{A} topical review},\ }\href
  {https://doi.org/10.1088/1751-8113/49/14/143001} {\bibfield  {journal}
  {\bibinfo  {journal} {J. Phys. A Math. Theor.}\ }\textbf {\bibinfo {volume}
  {49}},\ \bibinfo {pages} {143001} (\bibinfo {year} {2016})}\BibitemShut
  {NoStop}%
\bibitem [{\citenamefont {Binder}\ \emph {et~al.}(2018)\citenamefont {Binder},
  \citenamefont {Correa}, \citenamefont {Gogolin}, \citenamefont {Anders},\
  and\ \citenamefont {Adesso}}]{binder2018thermodynamics}%
  \BibitemOpen
  \bibinfo {editor} {\bibfnamefont {F.~C.}\ \bibnamefont {Binder}}, \bibinfo
  {editor} {\bibfnamefont {L.~A.}\ \bibnamefont {Correa}}, \bibinfo {editor}
  {\bibfnamefont {C.}~\bibnamefont {Gogolin}}, \bibinfo {editor} {\bibfnamefont
  {J.}~\bibnamefont {Anders}},\ and\ \bibinfo {editor} {\bibfnamefont
  {G.}~\bibnamefont {Adesso}},\ eds.,\ \href
  {https://doi.org/10.1007/978-3-319-99046-0} {\emph {\bibinfo {title}
  {Thermodynamics in the Quantum Regime: Fundamental Aspects and New
  Directions}}},\ \bibinfo {series} {Fundamental Theories of Physics}, Vol.\
  \bibinfo {volume} {195}\ (\bibinfo  {publisher} {Springer},\ \bibinfo {year}
  {2018})\BibitemShut {NoStop}%
\bibitem [{\citenamefont {Alicki}\ and\ \citenamefont
  {Kosloff}(2018)}]{alicki2019introduction}%
  \BibitemOpen
  \bibfield  {author} {\bibinfo {author} {\bibfnamefont {R.}~\bibnamefont
  {Alicki}}\ and\ \bibinfo {author} {\bibfnamefont {R.}~\bibnamefont
  {Kosloff}},\ }\bibfield  {title} {\bibinfo {title} {Introduction to quantum
  thermodynamics: History and prospects},\ }in\ \href
  {https://doi.org/10.1007/978-3-319-99046-0_1} {\emph {\bibinfo {booktitle}
  {Thermodynamics in the Quantum Regime: Fundamental Aspects and New
  Directions}}},\ \bibinfo {series} {Fundamental Theories of Physics}, Vol.\
  \bibinfo {volume} {195},\ \bibinfo {editor} {edited by\ \bibinfo {editor}
  {\bibfnamefont {F.~C.}\ \bibnamefont {Binder}}, \bibinfo {editor}
  {\bibfnamefont {L.~A.}\ \bibnamefont {Correa}}, \bibinfo {editor}
  {\bibfnamefont {C.}~\bibnamefont {Gogolin}}, \bibinfo {editor} {\bibfnamefont
  {J.}~\bibnamefont {Anders}},\ and\ \bibinfo {editor} {\bibfnamefont
  {G.}~\bibnamefont {Adesso}}}\ (\bibinfo  {publisher} {Springer},\ \bibinfo
  {year} {2018})\ pp.\ \bibinfo {pages} {1--33}\BibitemShut {NoStop}%
\bibitem [{\citenamefont {Campbell}\ \emph {et~al.}(2026)\citenamefont
  {Campbell}, \citenamefont {d'Amico}, \citenamefont {Ciampini}, \citenamefont
  {Anders}, \citenamefont {Ares}, \citenamefont {Artini}, \citenamefont
  {Auff{\`e}ves}, \citenamefont {Oftelie}, \citenamefont {Bettman},
  \citenamefont {Bonan{\c{c}}a} \emph {et~al.}}]{campbell2026roadmap}%
  \BibitemOpen
  \bibfield  {author} {\bibinfo {author} {\bibfnamefont {S.}~\bibnamefont
  {Campbell}}, \bibinfo {author} {\bibfnamefont {I.}~\bibnamefont {d'Amico}},
  \bibinfo {author} {\bibfnamefont {M.~A.}\ \bibnamefont {Ciampini}}, \bibinfo
  {author} {\bibfnamefont {J.}~\bibnamefont {Anders}}, \bibinfo {author}
  {\bibfnamefont {N.}~\bibnamefont {Ares}}, \bibinfo {author} {\bibfnamefont
  {S.}~\bibnamefont {Artini}}, \bibinfo {author} {\bibfnamefont
  {A.}~\bibnamefont {Auff{\`e}ves}}, \bibinfo {author} {\bibfnamefont {L.~B.}\
  \bibnamefont {Oftelie}}, \bibinfo {author} {\bibfnamefont {L.}~\bibnamefont
  {Bettman}}, \bibinfo {author} {\bibfnamefont {M.~V.}\ \bibnamefont
  {Bonan{\c{c}}a}}, \emph {et~al.},\ }\bibfield  {title} {\bibinfo {title}
  {Roadmap on quantum thermodynamics},\ }\href
  {https://doi.org/10.1088/2058-9565/ae1e27} {\bibfield  {journal} {\bibinfo
  {journal} {Quantum Sci. Technol.}\ }\textbf {\bibinfo {volume} {11}},\
  \bibinfo {pages} {012501} (\bibinfo {year} {2026})}\BibitemShut {NoStop}%
\bibitem [{\citenamefont {Alicki}\ and\ \citenamefont
  {Fannes}(2013)}]{alicki2013entanglement}%
  \BibitemOpen
  \bibfield  {author} {\bibinfo {author} {\bibfnamefont {R.}~\bibnamefont
  {Alicki}}\ and\ \bibinfo {author} {\bibfnamefont {M.}~\bibnamefont
  {Fannes}},\ }\bibfield  {title} {\bibinfo {title} {Entanglement boost for
  extractable work from ensembles of quantum batteries},\ }\href
  {https://doi.org/10.1103/PhysRevE.87.042123} {\bibfield  {journal} {\bibinfo
  {journal} {Phys. Rev. E}\ }\textbf {\bibinfo {volume} {87}},\ \bibinfo
  {pages} {042123} (\bibinfo {year} {2013})}\BibitemShut {NoStop}%
\bibitem [{\citenamefont {Hovhannisyan}\ \emph {et~al.}(2013)\citenamefont
  {Hovhannisyan}, \citenamefont {Perarnau-Llobet}, \citenamefont {Huber},\ and\
  \citenamefont {Ac\'{\i}n}}]{hovhannisyan2013entanglement}%
  \BibitemOpen
  \bibfield  {author} {\bibinfo {author} {\bibfnamefont {K.~V.}\ \bibnamefont
  {Hovhannisyan}}, \bibinfo {author} {\bibfnamefont {M.}~\bibnamefont
  {Perarnau-Llobet}}, \bibinfo {author} {\bibfnamefont {M.}~\bibnamefont
  {Huber}},\ and\ \bibinfo {author} {\bibfnamefont {A.}~\bibnamefont
  {Ac\'{\i}n}},\ }\bibfield  {title} {\bibinfo {title} {Entanglement generation
  is not necessary for optimal work extraction},\ }\href
  {https://doi.org/10.1103/PhysRevLett.111.240401} {\bibfield  {journal}
  {\bibinfo  {journal} {Phys. Rev. Lett.}\ }\textbf {\bibinfo {volume} {111}},\
  \bibinfo {pages} {240401} (\bibinfo {year} {2013})}\BibitemShut {NoStop}%
\bibitem [{\citenamefont {Binder}\ \emph
  {et~al.}(2015{\natexlab{a}})\citenamefont {Binder}, \citenamefont
  {Vinjanampathy}, \citenamefont {Modi},\ and\ \citenamefont
  {Goold}}]{binder2015operational}%
  \BibitemOpen
  \bibfield  {author} {\bibinfo {author} {\bibfnamefont {F.~C.}\ \bibnamefont
  {Binder}}, \bibinfo {author} {\bibfnamefont {S.}~\bibnamefont
  {Vinjanampathy}}, \bibinfo {author} {\bibfnamefont {K.}~\bibnamefont
  {Modi}},\ and\ \bibinfo {author} {\bibfnamefont {J.}~\bibnamefont {Goold}},\
  }\bibfield  {title} {\bibinfo {title} {Quantum thermodynamics of general
  quantum processes},\ }\href {https://doi.org/10.1103/PhysRevE.91.032119}
  {\bibfield  {journal} {\bibinfo  {journal} {Phys. Rev. E}\ }\textbf {\bibinfo
  {volume} {91}},\ \bibinfo {pages} {032119} (\bibinfo {year}
  {2015}{\natexlab{a}})}\BibitemShut {NoStop}%
\bibitem [{\citenamefont {Binder}\ \emph
  {et~al.}(2015{\natexlab{b}})\citenamefont {Binder}, \citenamefont
  {Vinjanampathy}, \citenamefont {Modi},\ and\ \citenamefont
  {Goold}}]{binder2015quantacell}%
  \BibitemOpen
  \bibfield  {author} {\bibinfo {author} {\bibfnamefont {F.~C.}\ \bibnamefont
  {Binder}}, \bibinfo {author} {\bibfnamefont {S.}~\bibnamefont
  {Vinjanampathy}}, \bibinfo {author} {\bibfnamefont {K.}~\bibnamefont
  {Modi}},\ and\ \bibinfo {author} {\bibfnamefont {J.}~\bibnamefont {Goold}},\
  }\bibfield  {title} {\bibinfo {title} {Quantacell: \text{Powerful} charging
  of quantum batteries},\ }\href
  {https://doi.org/10.1088/1367-2630/17/7/075015} {\bibfield  {journal}
  {\bibinfo  {journal} {New J. Phys.}\ }\textbf {\bibinfo {volume} {17}},\
  \bibinfo {pages} {075015} (\bibinfo {year} {2015}{\natexlab{b}})}\BibitemShut
  {NoStop}%
\bibitem [{\citenamefont {Campaioli}\ \emph {et~al.}(2017)\citenamefont
  {Campaioli}, \citenamefont {Pollock}, \citenamefont {Binder}, \citenamefont
  {C\'eleri}, \citenamefont {Goold}, \citenamefont {Vinjanampathy},\ and\
  \citenamefont {Modi}}]{campaioli2017enhancing}%
  \BibitemOpen
  \bibfield  {author} {\bibinfo {author} {\bibfnamefont {F.}~\bibnamefont
  {Campaioli}}, \bibinfo {author} {\bibfnamefont {F.~A.}\ \bibnamefont
  {Pollock}}, \bibinfo {author} {\bibfnamefont {F.~C.}\ \bibnamefont {Binder}},
  \bibinfo {author} {\bibfnamefont {L.}~\bibnamefont {C\'eleri}}, \bibinfo
  {author} {\bibfnamefont {J.}~\bibnamefont {Goold}}, \bibinfo {author}
  {\bibfnamefont {S.}~\bibnamefont {Vinjanampathy}},\ and\ \bibinfo {author}
  {\bibfnamefont {K.}~\bibnamefont {Modi}},\ }\bibfield  {title} {\bibinfo
  {title} {Enhancing the charging power of quantum batteries},\ }\href
  {https://doi.org/10.1103/PhysRevLett.118.150601} {\bibfield  {journal}
  {\bibinfo  {journal} {Phys. Rev. Lett.}\ }\textbf {\bibinfo {volume} {118}},\
  \bibinfo {pages} {150601} (\bibinfo {year} {2017})}\BibitemShut {NoStop}%
\bibitem [{\citenamefont {Ferraro}\ \emph {et~al.}(2018)\citenamefont
  {Ferraro}, \citenamefont {Campisi}, \citenamefont {Andolina}, \citenamefont
  {Pellegrini},\ and\ \citenamefont {Polini}}]{ferraro2018high}%
  \BibitemOpen
  \bibfield  {author} {\bibinfo {author} {\bibfnamefont {D.}~\bibnamefont
  {Ferraro}}, \bibinfo {author} {\bibfnamefont {M.}~\bibnamefont {Campisi}},
  \bibinfo {author} {\bibfnamefont {G.~M.}\ \bibnamefont {Andolina}}, \bibinfo
  {author} {\bibfnamefont {V.}~\bibnamefont {Pellegrini}},\ and\ \bibinfo
  {author} {\bibfnamefont {M.}~\bibnamefont {Polini}},\ }\bibfield  {title}
  {\bibinfo {title} {High-power collective charging of a solid-state quantum
  battery},\ }\href {https://doi.org/10.1103/PhysRevLett.120.117702} {\bibfield
   {journal} {\bibinfo  {journal} {Phys. Rev. Lett.}\ }\textbf {\bibinfo
  {volume} {120}},\ \bibinfo {pages} {117702} (\bibinfo {year}
  {2018})}\BibitemShut {NoStop}%
\bibitem [{\citenamefont {Le}\ \emph {et~al.}(2018)\citenamefont {Le},
  \citenamefont {Levinsen}, \citenamefont {Modi}, \citenamefont {Parish},\ and\
  \citenamefont {Pollock}}]{le2018spin}%
  \BibitemOpen
  \bibfield  {author} {\bibinfo {author} {\bibfnamefont {T.~P.}\ \bibnamefont
  {Le}}, \bibinfo {author} {\bibfnamefont {J.}~\bibnamefont {Levinsen}},
  \bibinfo {author} {\bibfnamefont {K.}~\bibnamefont {Modi}}, \bibinfo {author}
  {\bibfnamefont {M.~M.}\ \bibnamefont {Parish}},\ and\ \bibinfo {author}
  {\bibfnamefont {F.~A.}\ \bibnamefont {Pollock}},\ }\bibfield  {title}
  {\bibinfo {title} {Spin-chain model of a many-body quantum battery},\ }\href
  {https://doi.org/10.1103/PhysRevA.97.022106} {\bibfield  {journal} {\bibinfo
  {journal} {Phys. Rev. A}\ }\textbf {\bibinfo {volume} {97}},\ \bibinfo
  {pages} {022106} (\bibinfo {year} {2018})}\BibitemShut {NoStop}%
\bibitem [{\citenamefont {Henao}\ and\ \citenamefont
  {Serra}(2018)}]{henao2018role}%
  \BibitemOpen
  \bibfield  {author} {\bibinfo {author} {\bibfnamefont {I.}~\bibnamefont
  {Henao}}\ and\ \bibinfo {author} {\bibfnamefont {R.~M.}\ \bibnamefont
  {Serra}},\ }\bibfield  {title} {\bibinfo {title} {Role of quantum coherence
  in the thermodynamics of energy transfer},\ }\href
  {https://doi.org/10.1103/PhysRevE.97.062105} {\bibfield  {journal} {\bibinfo
  {journal} {Phys. Rev. E}\ }\textbf {\bibinfo {volume} {97}},\ \bibinfo
  {pages} {062105} (\bibinfo {year} {2018})}\BibitemShut {NoStop}%
\bibitem [{\citenamefont {Andolina}\ \emph {et~al.}(2019)\citenamefont
  {Andolina}, \citenamefont {Keck}, \citenamefont {Mari}, \citenamefont
  {Campisi}, \citenamefont {Giovannetti},\ and\ \citenamefont
  {Polini}}]{andolina2019extractable}%
  \BibitemOpen
  \bibfield  {author} {\bibinfo {author} {\bibfnamefont {G.~M.}\ \bibnamefont
  {Andolina}}, \bibinfo {author} {\bibfnamefont {M.}~\bibnamefont {Keck}},
  \bibinfo {author} {\bibfnamefont {A.}~\bibnamefont {Mari}}, \bibinfo {author}
  {\bibfnamefont {M.}~\bibnamefont {Campisi}}, \bibinfo {author} {\bibfnamefont
  {V.}~\bibnamefont {Giovannetti}},\ and\ \bibinfo {author} {\bibfnamefont
  {M.}~\bibnamefont {Polini}},\ }\bibfield  {title} {\bibinfo {title}
  {Extractable work, the role of correlations, and asymptotic freedom in
  quantum batteries},\ }\href {https://doi.org/10.1103/PhysRevLett.122.047702}
  {\bibfield  {journal} {\bibinfo  {journal} {Phys. Rev. Lett.}\ }\textbf
  {\bibinfo {volume} {122}},\ \bibinfo {pages} {047702} (\bibinfo {year}
  {2019})}\BibitemShut {NoStop}%
\bibitem [{\citenamefont {Barra}(2019)}]{barra2019dissipative}%
  \BibitemOpen
  \bibfield  {author} {\bibinfo {author} {\bibfnamefont {F.}~\bibnamefont
  {Barra}},\ }\bibfield  {title} {\bibinfo {title} {Dissipative charging of a
  quantum battery},\ }\href {https://doi.org/10.1103/PhysRevLett.122.210601}
  {\bibfield  {journal} {\bibinfo  {journal} {Phys. Rev. Lett.}\ }\textbf
  {\bibinfo {volume} {122}},\ \bibinfo {pages} {210601} (\bibinfo {year}
  {2019})}\BibitemShut {NoStop}%
\bibitem [{\citenamefont {Juli{\`a}-Farr{\'e}}\ \emph
  {et~al.}(2020)\citenamefont {Juli{\`a}-Farr{\'e}}, \citenamefont {Salamon},
  \citenamefont {Riera}, \citenamefont {Bera},\ and\ \citenamefont
  {Lewenstein}}]{julia2020bounds}%
  \BibitemOpen
  \bibfield  {author} {\bibinfo {author} {\bibfnamefont {S.}~\bibnamefont
  {Juli{\`a}-Farr{\'e}}}, \bibinfo {author} {\bibfnamefont {T.}~\bibnamefont
  {Salamon}}, \bibinfo {author} {\bibfnamefont {A.}~\bibnamefont {Riera}},
  \bibinfo {author} {\bibfnamefont {M.~N.}\ \bibnamefont {Bera}},\ and\
  \bibinfo {author} {\bibfnamefont {M.}~\bibnamefont {Lewenstein}},\ }\bibfield
   {title} {\bibinfo {title} {Bounds on the capacity and power of quantum
  batteries},\ }\href {https://doi.org/10.1103/PhysRevResearch.2.023113}
  {\bibfield  {journal} {\bibinfo  {journal} {Phys. Rev. Res.}\ }\textbf
  {\bibinfo {volume} {2}},\ \bibinfo {pages} {023113} (\bibinfo {year}
  {2020})}\BibitemShut {NoStop}%
\bibitem [{\citenamefont {Gyhm}\ \emph {et~al.}(2022)\citenamefont {Gyhm},
  \citenamefont {{\v{S}}afr{\'a}nek},\ and\ \citenamefont
  {Rosa}}]{gyhm2022quantum}%
  \BibitemOpen
  \bibfield  {author} {\bibinfo {author} {\bibfnamefont {J.-Y.}\ \bibnamefont
  {Gyhm}}, \bibinfo {author} {\bibfnamefont {D.}~\bibnamefont
  {{\v{S}}afr{\'a}nek}},\ and\ \bibinfo {author} {\bibfnamefont
  {D.}~\bibnamefont {Rosa}},\ }\bibfield  {title} {\bibinfo {title} {Quantum
  charging advantage cannot be extensive without global operations},\ }\href
  {https://doi.org/10.1103/PhysRevLett.128.140501} {\bibfield  {journal}
  {\bibinfo  {journal} {Phys. Rev. Lett.}\ }\textbf {\bibinfo {volume} {128}},\
  \bibinfo {pages} {140501} (\bibinfo {year} {2022})}\BibitemShut {NoStop}%
\bibitem [{\citenamefont {Quach}\ \emph {et~al.}(2023)\citenamefont {Quach},
  \citenamefont {Cerullo},\ and\ \citenamefont {Virgili}}]{quach2023quantum}%
  \BibitemOpen
  \bibfield  {author} {\bibinfo {author} {\bibfnamefont {J.~Q.}\ \bibnamefont
  {Quach}}, \bibinfo {author} {\bibfnamefont {G.}~\bibnamefont {Cerullo}},\
  and\ \bibinfo {author} {\bibfnamefont {T.}~\bibnamefont {Virgili}},\
  }\bibfield  {title} {\bibinfo {title} {Quantum batteries: \text{The} future
  of energy storage?},\ }\href {https://doi.org/10.1016/j.joule.2023.09.003}
  {\bibfield  {journal} {\bibinfo  {journal} {Joule}\ }\textbf {\bibinfo
  {volume} {7}},\ \bibinfo {pages} {2195} (\bibinfo {year} {2023})}\BibitemShut
  {NoStop}%
\bibitem [{\citenamefont {Ahmadi}\ \emph {et~al.}(2024)\citenamefont {Ahmadi},
  \citenamefont {Mazurek}, \citenamefont {Horodecki},\ and\ \citenamefont
  {Barzanjeh}}]{ahmadi2024nonreciprocal}%
  \BibitemOpen
  \bibfield  {author} {\bibinfo {author} {\bibfnamefont {B.}~\bibnamefont
  {Ahmadi}}, \bibinfo {author} {\bibfnamefont {P.}~\bibnamefont {Mazurek}},
  \bibinfo {author} {\bibfnamefont {P.}~\bibnamefont {Horodecki}},\ and\
  \bibinfo {author} {\bibfnamefont {S.}~\bibnamefont {Barzanjeh}},\ }\bibfield
  {title} {\bibinfo {title} {Nonreciprocal quantum batteries},\ }\href
  {https://doi.org/10.1103/PhysRevLett.132.210402} {\bibfield  {journal}
  {\bibinfo  {journal} {Phys. Rev. Lett.}\ }\textbf {\bibinfo {volume} {132}},\
  \bibinfo {pages} {210402} (\bibinfo {year} {2024})}\BibitemShut {NoStop}%
\bibitem [{\citenamefont {Lu}\ \emph {et~al.}(2025)\citenamefont {Lu},
  \citenamefont {Tian}, \citenamefont {L{\"u}},\ and\ \citenamefont
  {Shang}}]{lu2025topological}%
  \BibitemOpen
  \bibfield  {author} {\bibinfo {author} {\bibfnamefont {Z.-G.}\ \bibnamefont
  {Lu}}, \bibinfo {author} {\bibfnamefont {G.}~\bibnamefont {Tian}}, \bibinfo
  {author} {\bibfnamefont {X.-Y.}\ \bibnamefont {L{\"u}}},\ and\ \bibinfo
  {author} {\bibfnamefont {C.}~\bibnamefont {Shang}},\ }\bibfield  {title}
  {\bibinfo {title} {Topological quantum batteries},\ }\href
  {https://doi.org/10.1103/PhysRevLett.134.180401} {\bibfield  {journal}
  {\bibinfo  {journal} {Phys. Rev. Lett.}\ }\textbf {\bibinfo {volume} {134}},\
  \bibinfo {pages} {180401} (\bibinfo {year} {2025})}\BibitemShut {NoStop}%
\bibitem [{\citenamefont {Campaioli}\ \emph {et~al.}(2024)\citenamefont
  {Campaioli}, \citenamefont {Gherardini}, \citenamefont {Quach}, \citenamefont
  {Polini},\ and\ \citenamefont {Andolina}}]{campaioli2024colloquium}%
  \BibitemOpen
  \bibfield  {author} {\bibinfo {author} {\bibfnamefont {F.}~\bibnamefont
  {Campaioli}}, \bibinfo {author} {\bibfnamefont {S.}~\bibnamefont
  {Gherardini}}, \bibinfo {author} {\bibfnamefont {J.~Q.}\ \bibnamefont
  {Quach}}, \bibinfo {author} {\bibfnamefont {M.}~\bibnamefont {Polini}},\ and\
  \bibinfo {author} {\bibfnamefont {G.~M.}\ \bibnamefont {Andolina}},\
  }\bibfield  {title} {\bibinfo {title} {Colloquium: \text{Quantum}
  batteries},\ }\href {https://doi.org/10.1103/RevModPhys.96.031001} {\bibfield
   {journal} {\bibinfo  {journal} {Rev. Mod. Phys.}\ }\textbf {\bibinfo
  {volume} {96}},\ \bibinfo {pages} {031001} (\bibinfo {year}
  {2024})}\BibitemShut {NoStop}%
\bibitem [{\citenamefont {Kamin}\ \emph {et~al.}(2020)\citenamefont {Kamin},
  \citenamefont {Tabesh}, \citenamefont {Salimi},\ and\ \citenamefont
  {Santos}}]{kamin2020entanglement}%
  \BibitemOpen
  \bibfield  {author} {\bibinfo {author} {\bibfnamefont {F.~H.}\ \bibnamefont
  {Kamin}}, \bibinfo {author} {\bibfnamefont {F.~T.}\ \bibnamefont {Tabesh}},
  \bibinfo {author} {\bibfnamefont {S.}~\bibnamefont {Salimi}},\ and\ \bibinfo
  {author} {\bibfnamefont {A.~C.}\ \bibnamefont {Santos}},\ }\bibfield  {title}
  {\bibinfo {title} {Entanglement, coherence, and charging process of quantum
  batteries},\ }\href {https://doi.org/10.1103/PhysRevE.102.052109} {\bibfield
  {journal} {\bibinfo  {journal} {Phys. Rev. E}\ }\textbf {\bibinfo {volume}
  {102}},\ \bibinfo {pages} {052109} (\bibinfo {year} {2020})}\BibitemShut
  {NoStop}%
\bibitem [{\citenamefont {Shi}\ \emph {et~al.}(2022)\citenamefont {Shi},
  \citenamefont {Ding}, \citenamefont {Wan}, \citenamefont {Wang},\ and\
  \citenamefont {Yang}}]{shi2022entanglement}%
  \BibitemOpen
  \bibfield  {author} {\bibinfo {author} {\bibfnamefont {H.-L.}\ \bibnamefont
  {Shi}}, \bibinfo {author} {\bibfnamefont {S.}~\bibnamefont {Ding}}, \bibinfo
  {author} {\bibfnamefont {Q.-K.}\ \bibnamefont {Wan}}, \bibinfo {author}
  {\bibfnamefont {X.-H.}\ \bibnamefont {Wang}},\ and\ \bibinfo {author}
  {\bibfnamefont {W.-L.}\ \bibnamefont {Yang}},\ }\bibfield  {title} {\bibinfo
  {title} {Entanglement, coherence, and extractable work in quantum
  batteries},\ }\href {https://doi.org/10.1103/PhysRevLett.129.130602}
  {\bibfield  {journal} {\bibinfo  {journal} {Phys. Rev. Lett.}\ }\textbf
  {\bibinfo {volume} {129}},\ \bibinfo {pages} {130602} (\bibinfo {year}
  {2022})}\BibitemShut {NoStop}%
\bibitem [{\citenamefont {Gyhm}\ and\ \citenamefont
  {Fischer}(2024)}]{gyhm2024beneficial}%
  \BibitemOpen
  \bibfield  {author} {\bibinfo {author} {\bibfnamefont {J.-Y.}\ \bibnamefont
  {Gyhm}}\ and\ \bibinfo {author} {\bibfnamefont {U.~R.}\ \bibnamefont
  {Fischer}},\ }\bibfield  {title} {\bibinfo {title} {Beneficial and
  detrimental entanglement for quantum battery charging},\ }\href
  {https://doi.org/10.1116/5.0184903} {\bibfield  {journal} {\bibinfo
  {journal} {AVS Quantum Sci.}\ }\textbf {\bibinfo {volume} {6}},\ \bibinfo
  {pages} {012001} (\bibinfo {year} {2024})}\BibitemShut {NoStop}%
\bibitem [{\citenamefont {Santos}(2021)}]{santos2021quantum}%
  \BibitemOpen
  \bibfield  {author} {\bibinfo {author} {\bibfnamefont {A.~C.}\ \bibnamefont
  {Santos}},\ }\bibfield  {title} {\bibinfo {title} {Quantum advantage of
  two-level batteries in the self-discharging process},\ }\href
  {https://doi.org/10.1103/PhysRevE.103.042118} {\bibfield  {journal} {\bibinfo
   {journal} {Phys. Rev. E}\ }\textbf {\bibinfo {volume} {103}},\ \bibinfo
  {pages} {042118} (\bibinfo {year} {2021})}\BibitemShut {NoStop}%
\bibitem [{\citenamefont {Mohan}\ and\ \citenamefont
  {Pati}(2021)}]{mohan2021reverse}%
  \BibitemOpen
  \bibfield  {author} {\bibinfo {author} {\bibfnamefont {B.}~\bibnamefont
  {Mohan}}\ and\ \bibinfo {author} {\bibfnamefont {A.~K.}\ \bibnamefont
  {Pati}},\ }\bibfield  {title} {\bibinfo {title} {Reverse quantum speed limit:
  \text{How} slowly a quantum battery can discharge},\ }\href
  {https://doi.org/10.1103/PhysRevA.104.042209} {\bibfield  {journal} {\bibinfo
   {journal} {Phys. Rev. A}\ }\textbf {\bibinfo {volume} {104}},\ \bibinfo
  {pages} {042209} (\bibinfo {year} {2021})}\BibitemShut {NoStop}%
\bibitem [{\citenamefont {Arjmandi}\ \emph {et~al.}(2022)\citenamefont
  {Arjmandi}, \citenamefont {Mohammadi},\ and\ \citenamefont
  {Santos}}]{arjmandi2022enhancing}%
  \BibitemOpen
  \bibfield  {author} {\bibinfo {author} {\bibfnamefont {M.~B.}\ \bibnamefont
  {Arjmandi}}, \bibinfo {author} {\bibfnamefont {H.}~\bibnamefont
  {Mohammadi}},\ and\ \bibinfo {author} {\bibfnamefont {A.~C.}\ \bibnamefont
  {Santos}},\ }\bibfield  {title} {\bibinfo {title} {Enhancing self-discharging
  process with disordered quantum batteries},\ }\href
  {https://doi.org/10.1103/PhysRevE.105.054115} {\bibfield  {journal} {\bibinfo
   {journal} {Phys. Rev. E}\ }\textbf {\bibinfo {volume} {105}},\ \bibinfo
  {pages} {054115} (\bibinfo {year} {2022})}\BibitemShut {NoStop}%
\bibitem [{\citenamefont {Xu}\ \emph {et~al.}(2023)\citenamefont {Xu},
  \citenamefont {Zhu}, \citenamefont {Zhu}, \citenamefont {Zhang},\ and\
  \citenamefont {Liu}}]{xu2023charging}%
  \BibitemOpen
  \bibfield  {author} {\bibinfo {author} {\bibfnamefont {K.}~\bibnamefont
  {Xu}}, \bibinfo {author} {\bibfnamefont {H.-J.}\ \bibnamefont {Zhu}},
  \bibinfo {author} {\bibfnamefont {H.}~\bibnamefont {Zhu}}, \bibinfo {author}
  {\bibfnamefont {G.-F.}\ \bibnamefont {Zhang}},\ and\ \bibinfo {author}
  {\bibfnamefont {W.-M.}\ \bibnamefont {Liu}},\ }\bibfield  {title} {\bibinfo
  {title} {Charging and self-discharging process of a quantum battery in
  composite environments},\ }\href {https://doi.org/10.1007/s11467-022-1230-x}
  {\bibfield  {journal} {\bibinfo  {journal} {Front. Phys.}\ }\textbf {\bibinfo
  {volume} {18}},\ \bibinfo {pages} {31301} (\bibinfo {year}
  {2023})}\BibitemShut {NoStop}%
\bibitem [{\citenamefont {Song}\ \emph {et~al.}(2025)\citenamefont {Song},
  \citenamefont {Wang}, \citenamefont {Zhou}, \citenamefont {Yang},\ and\
  \citenamefont {An}}]{song2025self}%
  \BibitemOpen
  \bibfield  {author} {\bibinfo {author} {\bibfnamefont {W.-L.}\ \bibnamefont
  {Song}}, \bibinfo {author} {\bibfnamefont {J.-L.}\ \bibnamefont {Wang}},
  \bibinfo {author} {\bibfnamefont {B.}~\bibnamefont {Zhou}}, \bibinfo {author}
  {\bibfnamefont {W.-L.}\ \bibnamefont {Yang}},\ and\ \bibinfo {author}
  {\bibfnamefont {J.-H.}\ \bibnamefont {An}},\ }\bibfield  {title} {\bibinfo
  {title} {Self-discharging mitigated quantum battery},\ }\href
  {https://doi.org/10.1103/d9k1-75d4} {\bibfield  {journal} {\bibinfo
  {journal} {Phys. Rev. Lett.}\ }\textbf {\bibinfo {volume} {135}},\ \bibinfo
  {pages} {020405} (\bibinfo {year} {2025})}\BibitemShut {NoStop}%
\bibitem [{\citenamefont {Andolina}\ \emph {et~al.}(2018)\citenamefont
  {Andolina}, \citenamefont {Farina}, \citenamefont {Mari}, \citenamefont
  {Pellegrini}, \citenamefont {Giovannetti},\ and\ \citenamefont
  {Polini}}]{andolina2018charger}%
  \BibitemOpen
  \bibfield  {author} {\bibinfo {author} {\bibfnamefont {G.~M.}\ \bibnamefont
  {Andolina}}, \bibinfo {author} {\bibfnamefont {D.}~\bibnamefont {Farina}},
  \bibinfo {author} {\bibfnamefont {A.}~\bibnamefont {Mari}}, \bibinfo {author}
  {\bibfnamefont {V.}~\bibnamefont {Pellegrini}}, \bibinfo {author}
  {\bibfnamefont {V.}~\bibnamefont {Giovannetti}},\ and\ \bibinfo {author}
  {\bibfnamefont {M.}~\bibnamefont {Polini}},\ }\bibfield  {title} {\bibinfo
  {title} {Charger-mediated energy transfer in exactly solvable models for
  quantum batteries},\ }\href {https://doi.org/10.1103/PhysRevB.98.205423}
  {\bibfield  {journal} {\bibinfo  {journal} {Phys. Rev. B}\ }\textbf {\bibinfo
  {volume} {98}},\ \bibinfo {pages} {205423} (\bibinfo {year}
  {2018})}\BibitemShut {NoStop}%
\bibitem [{\citenamefont {Farina}\ \emph {et~al.}(2019)\citenamefont {Farina},
  \citenamefont {Andolina}, \citenamefont {Mari}, \citenamefont {Polini},\ and\
  \citenamefont {Giovannetti}}]{farina2019charger}%
  \BibitemOpen
  \bibfield  {author} {\bibinfo {author} {\bibfnamefont {D.}~\bibnamefont
  {Farina}}, \bibinfo {author} {\bibfnamefont {G.~M.}\ \bibnamefont
  {Andolina}}, \bibinfo {author} {\bibfnamefont {A.}~\bibnamefont {Mari}},
  \bibinfo {author} {\bibfnamefont {M.}~\bibnamefont {Polini}},\ and\ \bibinfo
  {author} {\bibfnamefont {V.}~\bibnamefont {Giovannetti}},\ }\bibfield
  {title} {\bibinfo {title} {Charger-mediated energy transfer for quantum
  batteries: \text{An} open-system approach},\ }\href
  {https://doi.org/10.1103/PhysRevB.99.035421} {\bibfield  {journal} {\bibinfo
  {journal} {Phys. Rev. B}\ }\textbf {\bibinfo {volume} {99}},\ \bibinfo
  {pages} {035421} (\bibinfo {year} {2019})}\BibitemShut {NoStop}%
\bibitem [{\citenamefont {Ukhtary}\ \emph {et~al.}(2023)\citenamefont
  {Ukhtary}, \citenamefont {Nugraha}, \citenamefont {Cahaya}, \citenamefont
  {Rusydi},\ and\ \citenamefont {Majidi}}]{ukhtary2023high}%
  \BibitemOpen
  \bibfield  {author} {\bibinfo {author} {\bibfnamefont {M.~S.}\ \bibnamefont
  {Ukhtary}}, \bibinfo {author} {\bibfnamefont {A.~R.~T.}\ \bibnamefont
  {Nugraha}}, \bibinfo {author} {\bibfnamefont {A.~B.}\ \bibnamefont {Cahaya}},
  \bibinfo {author} {\bibfnamefont {A.}~\bibnamefont {Rusydi}},\ and\ \bibinfo
  {author} {\bibfnamefont {M.~A.}\ \bibnamefont {Majidi}},\ }\bibfield  {title}
  {\bibinfo {title} {High-performance \text{Kerr} quantum battery},\ }\href
  {https://doi.org/10.1063/5.0156618} {\bibfield  {journal} {\bibinfo
  {journal} {Appl. Phys. Lett.}\ }\textbf {\bibinfo {volume} {123}},\ \bibinfo
  {pages} {034001} (\bibinfo {year} {2023})}\BibitemShut {NoStop}%
\bibitem [{\citenamefont {Downing}\ and\ \citenamefont
  {Ukhtary}(2023)}]{downing2023quantum}%
  \BibitemOpen
  \bibfield  {author} {\bibinfo {author} {\bibfnamefont {C.~A.}\ \bibnamefont
  {Downing}}\ and\ \bibinfo {author} {\bibfnamefont {M.~S.}\ \bibnamefont
  {Ukhtary}},\ }\bibfield  {title} {\bibinfo {title} {A quantum battery with
  quadratic driving},\ }\href {https://doi.org/10.1038/s42005-023-01439-y}
  {\bibfield  {journal} {\bibinfo  {journal} {Comm. Phys.}\ }\textbf {\bibinfo
  {volume} {6}},\ \bibinfo {pages} {322} (\bibinfo {year} {2023})}\BibitemShut
  {NoStop}%
\bibitem [{\citenamefont {Downing}\ and\ \citenamefont
  {Ukhtary}(2024{\natexlab{a}})}]{downing2024hyperbolic}%
  \BibitemOpen
  \bibfield  {author} {\bibinfo {author} {\bibfnamefont {C.~A.}\ \bibnamefont
  {Downing}}\ and\ \bibinfo {author} {\bibfnamefont {M.~S.}\ \bibnamefont
  {Ukhtary}},\ }\bibfield  {title} {\bibinfo {title} {Hyperbolic enhancement of
  a quantum battery},\ }\href {https://doi.org/10.1103/PhysRevA.109.052206}
  {\bibfield  {journal} {\bibinfo  {journal} {Phys. Rev. A}\ }\textbf {\bibinfo
  {volume} {109}},\ \bibinfo {pages} {052206} (\bibinfo {year}
  {2024}{\natexlab{a}})}\BibitemShut {NoStop}%
\bibitem [{\citenamefont {Downing}\ and\ \citenamefont
  {Ukhtary}(2024{\natexlab{b}})}]{downing2024energetics}%
  \BibitemOpen
  \bibfield  {author} {\bibinfo {author} {\bibfnamefont {C.~A.}\ \bibnamefont
  {Downing}}\ and\ \bibinfo {author} {\bibfnamefont {M.~S.}\ \bibnamefont
  {Ukhtary}},\ }\bibfield  {title} {\bibinfo {title} {Energetics of a pulsed
  quantum battery},\ }\href {https://doi.org/10.1209/0295-5075/ad2e79}
  {\bibfield  {journal} {\bibinfo  {journal} {Europhys. Lett.}\ }\textbf
  {\bibinfo {volume} {146}},\ \bibinfo {pages} {10001} (\bibinfo {year}
  {2024}{\natexlab{b}})}\BibitemShut {NoStop}%
\bibitem [{\citenamefont {Downing}\ and\ \citenamefont
  {Saroka}(2021)}]{downing2021exceptional}%
  \BibitemOpen
  \bibfield  {author} {\bibinfo {author} {\bibfnamefont {C.~A.}\ \bibnamefont
  {Downing}}\ and\ \bibinfo {author} {\bibfnamefont {V.~A.}\ \bibnamefont
  {Saroka}},\ }\bibfield  {title} {\bibinfo {title} {Exceptional points in
  oligomer chains},\ }\href {https://doi.org/10.1038/s42005-021-00757-3}
  {\bibfield  {journal} {\bibinfo  {journal} {Commun. Phys.}\ }\textbf
  {\bibinfo {volume} {4}},\ \bibinfo {pages} {254} (\bibinfo {year}
  {2021})}\BibitemShut {NoStop}%
\bibitem [{\citenamefont {Downing}\ and\ \citenamefont
  {Ukhtary}(2024{\natexlab{c}})}]{downing2024two}%
  \BibitemOpen
  \bibfield  {author} {\bibinfo {author} {\bibfnamefont {C.~A.}\ \bibnamefont
  {Downing}}\ and\ \bibinfo {author} {\bibfnamefont {M.~S.}\ \bibnamefont
  {Ukhtary}},\ }\bibfield  {title} {\bibinfo {title} {Two-photon charging of a
  quantum battery with a \text{Gaussian} pulse envelope},\ }\href
  {https://doi.org/10.1016/j.physleta.2024.129693} {\bibfield  {journal}
  {\bibinfo  {journal} {Phys. Lett. A}\ }\textbf {\bibinfo {volume} {518}},\
  \bibinfo {pages} {129693} (\bibinfo {year} {2024}{\natexlab{c}})}\BibitemShut
  {NoStop}%
\bibitem [{\citenamefont {Downing}\ and\ \citenamefont
  {Ukhtary}(2025)}]{downing2025energy}%
  \BibitemOpen
  \bibfield  {author} {\bibinfo {author} {\bibfnamefont {C.~A.}\ \bibnamefont
  {Downing}}\ and\ \bibinfo {author} {\bibfnamefont {M.~S.}\ \bibnamefont
  {Ukhtary}},\ }\bibfield  {title} {\bibinfo {title} {Energy storage in a
  continuous-variable quantum battery with nonlinear coupling},\ }\href
  {https://doi.org/10.1103/73zl-yn4h} {\bibfield  {journal} {\bibinfo
  {journal} {Phys. Rev. E}\ }\textbf {\bibinfo {volume} {112}},\ \bibinfo
  {pages} {044143} (\bibinfo {year} {2025})}\BibitemShut {NoStop}%
\bibitem [{\citenamefont {Rossini}\ \emph {et~al.}(2020)\citenamefont
  {Rossini}, \citenamefont {Andolina}, \citenamefont {Rosa}, \citenamefont
  {Carrega},\ and\ \citenamefont {Polini}}]{rossini2020quantum}%
  \BibitemOpen
  \bibfield  {author} {\bibinfo {author} {\bibfnamefont {D.}~\bibnamefont
  {Rossini}}, \bibinfo {author} {\bibfnamefont {G.~M.}\ \bibnamefont
  {Andolina}}, \bibinfo {author} {\bibfnamefont {D.}~\bibnamefont {Rosa}},
  \bibinfo {author} {\bibfnamefont {M.}~\bibnamefont {Carrega}},\ and\ \bibinfo
  {author} {\bibfnamefont {M.}~\bibnamefont {Polini}},\ }\bibfield  {title}
  {\bibinfo {title} {Quantum advantage in the charging process of
  \text{Sachdev-Ye-Kitaev} batteries},\ }\href
  {https://doi.org/10.1103/PhysRevLett.125.236402} {\bibfield  {journal}
  {\bibinfo  {journal} {Phys. Rev. Lett.}\ }\textbf {\bibinfo {volume} {125}},\
  \bibinfo {pages} {236402} (\bibinfo {year} {2020})}\BibitemShut {NoStop}%
\bibitem [{\citenamefont {Rosa}\ \emph {et~al.}(2020)\citenamefont {Rosa},
  \citenamefont {Rossini}, \citenamefont {Andolina}, \citenamefont {Polini},\
  and\ \citenamefont {Carrega}}]{rosa2020ultra}%
  \BibitemOpen
  \bibfield  {author} {\bibinfo {author} {\bibfnamefont {D.}~\bibnamefont
  {Rosa}}, \bibinfo {author} {\bibfnamefont {D.}~\bibnamefont {Rossini}},
  \bibinfo {author} {\bibfnamefont {G.~M.}\ \bibnamefont {Andolina}}, \bibinfo
  {author} {\bibfnamefont {M.}~\bibnamefont {Polini}},\ and\ \bibinfo {author}
  {\bibfnamefont {M.}~\bibnamefont {Carrega}},\ }\bibfield  {title} {\bibinfo
  {title} {Ultra-stable charging of fast-scrambling \text{SYK} quantum
  batteries},\ }\href {https://doi.org/10.1007/JHEP11(2020)067} {\bibfield
  {journal} {\bibinfo  {journal} {\text{J. High Energy Phys.}}\ }\textbf
  {\bibinfo {volume} {2020}},\ \bibinfo {pages} {1} (\bibinfo {year}
  {2020})}\BibitemShut {NoStop}%
\bibitem [{\citenamefont {Francica}(2024)}]{francica2024quantum}%
  \BibitemOpen
  \bibfield  {author} {\bibinfo {author} {\bibfnamefont {G.}~\bibnamefont
  {Francica}},\ }\bibfield  {title} {\bibinfo {title} {Quantum advantage in
  batteries for \text{Sachdev-Ye-Kitaev} interactions},\ }\href
  {https://doi.org/10.1103/PhysRevA.110.062209} {\bibfield  {journal} {\bibinfo
   {journal} {Phys. Rev. A}\ }\textbf {\bibinfo {volume} {110}},\ \bibinfo
  {pages} {062209} (\bibinfo {year} {2024})}\BibitemShut {NoStop}%
\bibitem [{\citenamefont {Romero}\ \emph {et~al.}(2025)\citenamefont {Romero},
  \citenamefont {Ding}, \citenamefont {Chen},\ and\ \citenamefont
  {Ban}}]{romero2025scrambling}%
  \BibitemOpen
  \bibfield  {author} {\bibinfo {author} {\bibfnamefont {S.~V.}\ \bibnamefont
  {Romero}}, \bibinfo {author} {\bibfnamefont {Y.}~\bibnamefont {Ding}},
  \bibinfo {author} {\bibfnamefont {X.}~\bibnamefont {Chen}},\ and\ \bibinfo
  {author} {\bibfnamefont {Y.}~\bibnamefont {Ban}},\ }\bibfield  {title}
  {\bibinfo {title} {Scrambling in the charging of quantum batteries},\ }\href
  {https://doi.org/10.1007/JHEP05(2025)021} {\bibfield  {journal} {\bibinfo
  {journal} {\text{J. High Energy Phys.}}\ }\textbf {\bibinfo {volume}
  {2025}},\ \bibinfo {pages} {1} (\bibinfo {year} {2025})}\BibitemShut
  {NoStop}%
\bibitem [{\citenamefont {Divi}\ \emph {et~al.}(2025)\citenamefont {Divi},
  \citenamefont {Murugan},\ and\ \citenamefont {Rosa}}]{divi2025sachdev}%
  \BibitemOpen
  \bibfield  {author} {\bibinfo {author} {\bibfnamefont {F.}~\bibnamefont
  {Divi}}, \bibinfo {author} {\bibfnamefont {J.}~\bibnamefont {Murugan}},\ and\
  \bibinfo {author} {\bibfnamefont {D.}~\bibnamefont {Rosa}},\ }\bibfield
  {title} {\bibinfo {title} {\text{Sachdev-Ye-Kitaev} charging advantage as a
  random walk on graphs},\ }\href {https://doi.org/10.1103/PhysRevB.111.075138}
  {\bibfield  {journal} {\bibinfo  {journal} {Phys. Rev. B}\ }\textbf {\bibinfo
  {volume} {111}},\ \bibinfo {pages} {075138} (\bibinfo {year}
  {2025})}\BibitemShut {NoStop}%
\bibitem [{\citenamefont {Crescente}\ \emph
  {et~al.}(2020{\natexlab{a}})\citenamefont {Crescente}, \citenamefont
  {Carrega}, \citenamefont {Sassetti},\ and\ \citenamefont
  {Ferraro}}]{crescente2020charging}%
  \BibitemOpen
  \bibfield  {author} {\bibinfo {author} {\bibfnamefont {A.}~\bibnamefont
  {Crescente}}, \bibinfo {author} {\bibfnamefont {M.}~\bibnamefont {Carrega}},
  \bibinfo {author} {\bibfnamefont {M.}~\bibnamefont {Sassetti}},\ and\
  \bibinfo {author} {\bibfnamefont {D.}~\bibnamefont {Ferraro}},\ }\bibfield
  {title} {\bibinfo {title} {Charging and energy fluctuations of a driven
  quantum battery},\ }\href {https://doi.org/10.1088/1367-2630/ab91fc}
  {\bibfield  {journal} {\bibinfo  {journal} {New J. Phys.}\ }\textbf {\bibinfo
  {volume} {22}},\ \bibinfo {pages} {063057} (\bibinfo {year}
  {2020}{\natexlab{a}})}\BibitemShut {NoStop}%
\bibitem [{\citenamefont {Crescente}\ \emph
  {et~al.}(2020{\natexlab{b}})\citenamefont {Crescente}, \citenamefont
  {Carrega}, \citenamefont {Sassetti},\ and\ \citenamefont
  {Ferraro}}]{crescente2020ultrafast}%
  \BibitemOpen
  \bibfield  {author} {\bibinfo {author} {\bibfnamefont {A.}~\bibnamefont
  {Crescente}}, \bibinfo {author} {\bibfnamefont {M.}~\bibnamefont {Carrega}},
  \bibinfo {author} {\bibfnamefont {M.}~\bibnamefont {Sassetti}},\ and\
  \bibinfo {author} {\bibfnamefont {D.}~\bibnamefont {Ferraro}},\ }\bibfield
  {title} {\bibinfo {title} {Ultrafast charging in a two-photon \text{Dicke}
  quantum battery},\ }\href {https://doi.org/10.1103/PhysRevB.102.245407}
  {\bibfield  {journal} {\bibinfo  {journal} {Phys. Rev. B}\ }\textbf {\bibinfo
  {volume} {102}},\ \bibinfo {pages} {245407} (\bibinfo {year}
  {2020}{\natexlab{b}})}\BibitemShut {NoStop}%
\bibitem [{\citenamefont {Dou}\ \emph {et~al.}(2022)\citenamefont {Dou},
  \citenamefont {Lu}, \citenamefont {Wang},\ and\ \citenamefont
  {Sun}}]{dou2022extended}%
  \BibitemOpen
  \bibfield  {author} {\bibinfo {author} {\bibfnamefont {F.-Q.}\ \bibnamefont
  {Dou}}, \bibinfo {author} {\bibfnamefont {Y.-Q.}\ \bibnamefont {Lu}},
  \bibinfo {author} {\bibfnamefont {Y.-J.}\ \bibnamefont {Wang}},\ and\
  \bibinfo {author} {\bibfnamefont {J.-A.}\ \bibnamefont {Sun}},\ }\bibfield
  {title} {\bibinfo {title} {Extended \text{Dicke} quantum battery with
  interatomic interactions and driving field},\ }\href
  {https://doi.org/10.1103/PhysRevB.105.115405} {\bibfield  {journal} {\bibinfo
   {journal} {Phys. Rev. B}\ }\textbf {\bibinfo {volume} {105}},\ \bibinfo
  {pages} {115405} (\bibinfo {year} {2022})}\BibitemShut {NoStop}%
\bibitem [{\citenamefont {Mazzoncini}\ \emph {et~al.}(2023)\citenamefont
  {Mazzoncini}, \citenamefont {Cavina}, \citenamefont {Andolina}, \citenamefont
  {Erdman},\ and\ \citenamefont {Giovannetti}}]{mazzoncini2023optimal}%
  \BibitemOpen
  \bibfield  {author} {\bibinfo {author} {\bibfnamefont {F.}~\bibnamefont
  {Mazzoncini}}, \bibinfo {author} {\bibfnamefont {V.}~\bibnamefont {Cavina}},
  \bibinfo {author} {\bibfnamefont {G.~M.}\ \bibnamefont {Andolina}}, \bibinfo
  {author} {\bibfnamefont {P.~A.}\ \bibnamefont {Erdman}},\ and\ \bibinfo
  {author} {\bibfnamefont {V.}~\bibnamefont {Giovannetti}},\ }\bibfield
  {title} {\bibinfo {title} {Optimal control methods for quantum batteries},\
  }\href {https://doi.org/10.1103/PhysRevA.107.032218} {\bibfield  {journal}
  {\bibinfo  {journal} {Phys. Rev. A}\ }\textbf {\bibinfo {volume} {107}},\
  \bibinfo {pages} {032218} (\bibinfo {year} {2023})}\BibitemShut {NoStop}%
\bibitem [{\citenamefont {Rodriguez}\ \emph {et~al.}(2024)\citenamefont
  {Rodriguez}, \citenamefont {Ahmadi}, \citenamefont {Su{\'a}rez},
  \citenamefont {Mazurek}, \citenamefont {Barzanjeh},\ and\ \citenamefont
  {Horodecki}}]{rodriguez2024optimal}%
  \BibitemOpen
  \bibfield  {author} {\bibinfo {author} {\bibfnamefont {R.~R.}\ \bibnamefont
  {Rodriguez}}, \bibinfo {author} {\bibfnamefont {B.}~\bibnamefont {Ahmadi}},
  \bibinfo {author} {\bibfnamefont {G.}~\bibnamefont {Su{\'a}rez}}, \bibinfo
  {author} {\bibfnamefont {P.}~\bibnamefont {Mazurek}}, \bibinfo {author}
  {\bibfnamefont {S.}~\bibnamefont {Barzanjeh}},\ and\ \bibinfo {author}
  {\bibfnamefont {P.}~\bibnamefont {Horodecki}},\ }\bibfield  {title} {\bibinfo
  {title} {Optimal quantum control of charging quantum batteries},\ }\href
  {https://doi.org/10.1088/1367-2630/ad3843} {\bibfield  {journal} {\bibinfo
  {journal} {New J. Phys.}\ }\textbf {\bibinfo {volume} {26}},\ \bibinfo
  {pages} {043004} (\bibinfo {year} {2024})}\BibitemShut {NoStop}%
\bibitem [{\citenamefont {Mitchison}\ \emph {et~al.}(2021)\citenamefont
  {Mitchison}, \citenamefont {Goold},\ and\ \citenamefont
  {Prior}}]{mitchison2021charging}%
  \BibitemOpen
  \bibfield  {author} {\bibinfo {author} {\bibfnamefont {M.~T.}\ \bibnamefont
  {Mitchison}}, \bibinfo {author} {\bibfnamefont {J.}~\bibnamefont {Goold}},\
  and\ \bibinfo {author} {\bibfnamefont {J.}~\bibnamefont {Prior}},\ }\bibfield
   {title} {\bibinfo {title} {Charging a quantum battery with linear feedback
  control},\ }\href {https://doi.org/10.22331/q-2021-07-13-500} {\bibfield
  {journal} {\bibinfo  {journal} {Quantum}\ }\textbf {\bibinfo {volume} {5}},\
  \bibinfo {pages} {500} (\bibinfo {year} {2021})}\BibitemShut {NoStop}%
\bibitem [{\citenamefont {Gorini}\ \emph {et~al.}(1976)\citenamefont {Gorini},
  \citenamefont {Kossakowski},\ and\ \citenamefont
  {Sudarshan}}]{gorini1976completely}%
  \BibitemOpen
  \bibfield  {author} {\bibinfo {author} {\bibfnamefont {V.}~\bibnamefont
  {Gorini}}, \bibinfo {author} {\bibfnamefont {A.}~\bibnamefont
  {Kossakowski}},\ and\ \bibinfo {author} {\bibfnamefont {E.~C.~G.}\
  \bibnamefont {Sudarshan}},\ }\bibfield  {title} {\bibinfo {title} {Completely
  positive dynamical semigroups of $n$-level systems},\ }\href
  {https://doi.org/10.1063/1.522979} {\bibfield  {journal} {\bibinfo  {journal}
  {J. Math. Phys.}\ }\textbf {\bibinfo {volume} {17}},\ \bibinfo {pages} {821}
  (\bibinfo {year} {1976})}\BibitemShut {NoStop}%
\bibitem [{\citenamefont {Lindblad}(1976)}]{lindblad1976onthegenerators}%
  \BibitemOpen
  \bibfield  {author} {\bibinfo {author} {\bibfnamefont {G.}~\bibnamefont
  {Lindblad}},\ }\bibfield  {title} {\bibinfo {title} {On the generators of
  quantum dynamical semigroups},\ }\href {https://doi.org/10.1007/BF01608499}
  {\bibfield  {journal} {\bibinfo  {journal} {Commun. Math. Phys.}\ }\textbf
  {\bibinfo {volume} {48}},\ \bibinfo {pages} {119} (\bibinfo {year}
  {1976})}\BibitemShut {NoStop}%
\bibitem [{\citenamefont {Ashida}\ \emph {et~al.}(2020)\citenamefont {Ashida},
  \citenamefont {Gong},\ and\ \citenamefont {Ueda}}]{ashida2020nonhermitian}%
  \BibitemOpen
  \bibfield  {author} {\bibinfo {author} {\bibfnamefont {Y.}~\bibnamefont
  {Ashida}}, \bibinfo {author} {\bibfnamefont {Z.}~\bibnamefont {Gong}},\ and\
  \bibinfo {author} {\bibfnamefont {M.}~\bibnamefont {Ueda}},\ }\bibfield
  {title} {\bibinfo {title} {\text{Non-Hermitian} physics},\ }\href
  {https://doi.org/10.1080/00018732.2021.1876991} {\bibfield  {journal}
  {\bibinfo  {journal} {Adv. Phys.}\ }\textbf {\bibinfo {volume} {69}},\
  \bibinfo {pages} {249} (\bibinfo {year} {2020})}\BibitemShut {NoStop}%
\bibitem [{\citenamefont {Wong}(1975)}]{wong75onthegeneralized}%
  \BibitemOpen
  \bibfield  {author} {\bibinfo {author} {\bibfnamefont {J.~S.~W.}\
  \bibnamefont {Wong}},\ }\bibfield  {title} {\bibinfo {title} {On the
  generalized \text{Emden–Fowler} equation},\ }\href
  {https://doi.org/10.1137/1017036} {\bibfield  {journal} {\bibinfo  {journal}
  {SIAM Rev.}\ }\textbf {\bibinfo {volume} {17}},\ \bibinfo {pages} {339}
  (\bibinfo {year} {1975})}\BibitemShut {NoStop}%
\bibitem [{\citenamefont {Quach}\ \emph {et~al.}(2022)\citenamefont {Quach},
  \citenamefont {McGhee}, \citenamefont {Ganzer}, \citenamefont {Rouse},
  \citenamefont {Lovett}, \citenamefont {Gauger}, \citenamefont {Keeling},
  \citenamefont {Cerullo}, \citenamefont {Lidzey},\ and\ \citenamefont
  {Virgili}}]{quach2022superabsorption}%
  \BibitemOpen
  \bibfield  {author} {\bibinfo {author} {\bibfnamefont {J.~Q.}\ \bibnamefont
  {Quach}}, \bibinfo {author} {\bibfnamefont {K.~E.}\ \bibnamefont {McGhee}},
  \bibinfo {author} {\bibfnamefont {L.}~\bibnamefont {Ganzer}}, \bibinfo
  {author} {\bibfnamefont {D.~M.}\ \bibnamefont {Rouse}}, \bibinfo {author}
  {\bibfnamefont {B.~W.}\ \bibnamefont {Lovett}}, \bibinfo {author}
  {\bibfnamefont {E.~M.}\ \bibnamefont {Gauger}}, \bibinfo {author}
  {\bibfnamefont {J.}~\bibnamefont {Keeling}}, \bibinfo {author} {\bibfnamefont
  {G.}~\bibnamefont {Cerullo}}, \bibinfo {author} {\bibfnamefont {D.~G.}\
  \bibnamefont {Lidzey}},\ and\ \bibinfo {author} {\bibfnamefont
  {T.}~\bibnamefont {Virgili}},\ }\bibfield  {title} {\bibinfo {title}
  {Superabsorption in an organic microcavity: \text{Toward} a quantum
  battery},\ }\href {https://doi.org/10.1126/sciadv.abk3160} {\bibfield
  {journal} {\bibinfo  {journal} {Sci. Adv.}\ }\textbf {\bibinfo {volume}
  {8}},\ \bibinfo {pages} {eabk3160} (\bibinfo {year} {2022})}\BibitemShut
  {NoStop}%
\bibitem [{\citenamefont {Hymas}\ \emph {et~al.}(2026)\citenamefont {Hymas},
  \citenamefont {Muir}, \citenamefont {Tibben}, \citenamefont {van Embden},
  \citenamefont {Hirai}, \citenamefont {Dunn}, \citenamefont {Gomez},
  \citenamefont {Hutchison}, \citenamefont {Smith},\ and\ \citenamefont
  {Quach}}]{Hymas2026SuperextensiveElectricalPower}%
  \BibitemOpen
  \bibfield  {author} {\bibinfo {author} {\bibfnamefont {K.}~\bibnamefont
  {Hymas}}, \bibinfo {author} {\bibfnamefont {J.~B.}\ \bibnamefont {Muir}},
  \bibinfo {author} {\bibfnamefont {D.}~\bibnamefont {Tibben}}, \bibinfo
  {author} {\bibfnamefont {J.}~\bibnamefont {van Embden}}, \bibinfo {author}
  {\bibfnamefont {T.}~\bibnamefont {Hirai}}, \bibinfo {author} {\bibfnamefont
  {C.~J.}\ \bibnamefont {Dunn}}, \bibinfo {author} {\bibfnamefont {D.~E.}\
  \bibnamefont {Gomez}}, \bibinfo {author} {\bibfnamefont {J.~A.}\ \bibnamefont
  {Hutchison}}, \bibinfo {author} {\bibfnamefont {T.~A.}\ \bibnamefont
  {Smith}},\ and\ \bibinfo {author} {\bibfnamefont {J.~Q.}\ \bibnamefont
  {Quach}},\ }\bibfield  {title} {\bibinfo {title} {Superextensive electrical
  power from a quantum battery},\ }\href
  {https://doi.org/10.1038/s41377-026-02240-6} {\bibfield  {journal} {\bibinfo
  {journal} {Light Sci. Appl.}\ }\textbf {\bibinfo {volume} {15}},\ \bibinfo
  {pages} {168} (\bibinfo {year} {2026})}\BibitemShut {NoStop}%
\bibitem [{\citenamefont {Lu}\ \emph {et~al.}(2021)\citenamefont {Lu},
  \citenamefont {Chen}, \citenamefont {Kuang},\ and\ \citenamefont
  {Wang}}]{lu2021optimal}%
  \BibitemOpen
  \bibfield  {author} {\bibinfo {author} {\bibfnamefont {W.}~\bibnamefont
  {Lu}}, \bibinfo {author} {\bibfnamefont {J.}~\bibnamefont {Chen}}, \bibinfo
  {author} {\bibfnamefont {L.-M.}\ \bibnamefont {Kuang}},\ and\ \bibinfo
  {author} {\bibfnamefont {X.}~\bibnamefont {Wang}},\ }\bibfield  {title}
  {\bibinfo {title} {Optimal state for a \text{Tavis-Cummings} quantum battery
  via the \text{Bethe} ansatz method},\ }\href
  {https://doi.org/10.1103/PhysRevA.104.043706} {\bibfield  {journal} {\bibinfo
   {journal} {Phys. Rev. A}\ }\textbf {\bibinfo {volume} {104}},\ \bibinfo
  {pages} {043706} (\bibinfo {year} {2021})}\BibitemShut {NoStop}%
\bibitem [{\citenamefont {Yang}\ \emph {et~al.}(2024)\citenamefont {Yang},
  \citenamefont {Shi}, \citenamefont {Wan}, \citenamefont {Zhang},
  \citenamefont {Wang},\ and\ \citenamefont {Yang}}]{yang2024optimal}%
  \BibitemOpen
  \bibfield  {author} {\bibinfo {author} {\bibfnamefont {H.-Y.}\ \bibnamefont
  {Yang}}, \bibinfo {author} {\bibfnamefont {H.-L.}\ \bibnamefont {Shi}},
  \bibinfo {author} {\bibfnamefont {Q.-K.}\ \bibnamefont {Wan}}, \bibinfo
  {author} {\bibfnamefont {K.}~\bibnamefont {Zhang}}, \bibinfo {author}
  {\bibfnamefont {X.-H.}\ \bibnamefont {Wang}},\ and\ \bibinfo {author}
  {\bibfnamefont {W.-L.}\ \bibnamefont {Yang}},\ }\bibfield  {title} {\bibinfo
  {title} {Optimal energy storage in the \text{Tavis-Cummings} quantum
  battery},\ }\href {https://doi.org/10.1103/PhysRevA.109.012204} {\bibfield
  {journal} {\bibinfo  {journal} {Phys. Rev. A}\ }\textbf {\bibinfo {volume}
  {109}},\ \bibinfo {pages} {012204} (\bibinfo {year} {2024})}\BibitemShut
  {NoStop}%
\bibitem [{\citenamefont {Canzio}\ \emph {et~al.}(2025)\citenamefont {Canzio},
  \citenamefont {Cavina}, \citenamefont {Polini},\ and\ \citenamefont
  {Giovannetti}}]{canzio2025single}%
  \BibitemOpen
  \bibfield  {author} {\bibinfo {author} {\bibfnamefont {A.}~\bibnamefont
  {Canzio}}, \bibinfo {author} {\bibfnamefont {V.}~\bibnamefont {Cavina}},
  \bibinfo {author} {\bibfnamefont {M.}~\bibnamefont {Polini}},\ and\ \bibinfo
  {author} {\bibfnamefont {V.}~\bibnamefont {Giovannetti}},\ }\bibfield
  {title} {\bibinfo {title} {Single-atom dissipation and dephasing in
  \text{Dicke and Tavis-Cummings} quantum batteries},\ }\href
  {https://doi.org/10.1103/PhysRevA.111.022222} {\bibfield  {journal} {\bibinfo
   {journal} {Phys. Rev. A}\ }\textbf {\bibinfo {volume} {111}},\ \bibinfo
  {pages} {022222} (\bibinfo {year} {2025})}\BibitemShut {NoStop}%
\bibitem [{\citenamefont {Dicke}(1954)}]{dicke1954coherence}%
  \BibitemOpen
  \bibfield  {author} {\bibinfo {author} {\bibfnamefont {R.~H.}\ \bibnamefont
  {Dicke}},\ }\bibfield  {title} {\bibinfo {title} {Coherence in spontaneous
  radiation processes},\ }\href {https://doi.org/10.1103/PhysRev.93.99}
  {\bibfield  {journal} {\bibinfo  {journal} {Phys. Rev.}\ }\textbf {\bibinfo
  {volume} {93}},\ \bibinfo {pages} {99} (\bibinfo {year} {1954})}\BibitemShut
  {NoStop}%
\bibitem [{\citenamefont {Bastidas}\ \emph {et~al.}(2012)\citenamefont
  {Bastidas}, \citenamefont {Emary}, \citenamefont {Regler},\ and\
  \citenamefont {Brandes}}]{bastidas2012nonequilibrium}%
  \BibitemOpen
  \bibfield  {author} {\bibinfo {author} {\bibfnamefont {V.~M.}\ \bibnamefont
  {Bastidas}}, \bibinfo {author} {\bibfnamefont {C.}~\bibnamefont {Emary}},
  \bibinfo {author} {\bibfnamefont {B.}~\bibnamefont {Regler}},\ and\ \bibinfo
  {author} {\bibfnamefont {T.}~\bibnamefont {Brandes}},\ }\bibfield  {title}
  {\bibinfo {title} {Nonequilibrium quantum phase transitions in the
  \text{Dicke} model},\ }\href {https://doi.org/10.1103/PhysRevLett.108.043003}
  {\bibfield  {journal} {\bibinfo  {journal} {Phys. Rev. Lett.}\ }\textbf
  {\bibinfo {volume} {108}},\ \bibinfo {pages} {043003} (\bibinfo {year}
  {2012})}\BibitemShut {NoStop}%
\bibitem [{\citenamefont {Jara~Jr}\ and\ \citenamefont
  {Cosme}(2024)}]{jara2024apparent}%
  \BibitemOpen
  \bibfield  {author} {\bibinfo {author} {\bibfnamefont {R.~D.}\ \bibnamefont
  {Jara~Jr}}\ and\ \bibinfo {author} {\bibfnamefont {J.~G.}\ \bibnamefont
  {Cosme}},\ }\bibfield  {title} {\bibinfo {title} {Apparent delay of the
  \text{Kibble-Zurek} mechanism in quenched open systems},\ }\href
  {https://doi.org/10.1103/PhysRevB.110.064317} {\bibfield  {journal} {\bibinfo
   {journal} {Phys. Rev. B}\ }\textbf {\bibinfo {volume} {110}},\ \bibinfo
  {pages} {064317} (\bibinfo {year} {2024})}\BibitemShut {NoStop}%
\bibitem [{\citenamefont {Kibble}(1976)}]{kibble1976topology}%
  \BibitemOpen
  \bibfield  {author} {\bibinfo {author} {\bibfnamefont {T.~W.~B.}\
  \bibnamefont {Kibble}},\ }\bibfield  {title} {\bibinfo {title} {Topology of
  cosmic domains and strings},\ }\href
  {https://doi.org/10.1088/0305-4470/9/8/029} {\bibfield  {journal} {\bibinfo
  {journal} {J. Phys. A Math. Gen.}\ }\textbf {\bibinfo {volume} {9}},\
  \bibinfo {pages} {1387} (\bibinfo {year} {1976})}\BibitemShut {NoStop}%
\bibitem [{\citenamefont {Kibble}(1980)}]{kibble1980some}%
  \BibitemOpen
  \bibfield  {author} {\bibinfo {author} {\bibfnamefont {T.~W.~B.}\
  \bibnamefont {Kibble}},\ }\bibfield  {title} {\bibinfo {title} {Some
  implications of a cosmological phase transition},\ }\href
  {https://doi.org/10.1016/0370-1573(80)90091-5} {\bibfield  {journal}
  {\bibinfo  {journal} {Phys. Rep.}\ }\textbf {\bibinfo {volume} {67}},\
  \bibinfo {pages} {183} (\bibinfo {year} {1980})}\BibitemShut {NoStop}%
\bibitem [{\citenamefont {Zurek}(1985)}]{zurek1985cosmological}%
  \BibitemOpen
  \bibfield  {author} {\bibinfo {author} {\bibfnamefont {W.~H.}\ \bibnamefont
  {Zurek}},\ }\bibfield  {title} {\bibinfo {title} {Cosmological experiments in
  superfluid helium?},\ }\href {https://doi.org/10.1038/317505a0} {\bibfield
  {journal} {\bibinfo  {journal} {Nature}\ }\textbf {\bibinfo {volume} {317}},\
  \bibinfo {pages} {505} (\bibinfo {year} {1985})}\BibitemShut {NoStop}%
\bibitem [{\citenamefont {Zurek}(1996)}]{zurek1996cosmological}%
  \BibitemOpen
  \bibfield  {author} {\bibinfo {author} {\bibfnamefont {W.~H.}\ \bibnamefont
  {Zurek}},\ }\bibfield  {title} {\bibinfo {title} {Cosmological experiments in
  condensed matter systems},\ }\href
  {https://doi.org/10.1016/S0370-1573(96)00009-9} {\bibfield  {journal}
  {\bibinfo  {journal} {Phys. Rep.}\ }\textbf {\bibinfo {volume} {276}},\
  \bibinfo {pages} {177} (\bibinfo {year} {1996})}\BibitemShut {NoStop}%
\bibitem [{\citenamefont {del Campo}\ and\ \citenamefont
  {Zurek}(2014)}]{campo2014universality}%
  \BibitemOpen
  \bibfield  {author} {\bibinfo {author} {\bibfnamefont {A.}~\bibnamefont {del
  Campo}}\ and\ \bibinfo {author} {\bibfnamefont {W.~H.}\ \bibnamefont
  {Zurek}},\ }\bibfield  {title} {\bibinfo {title} {Universality of phase
  transition dynamics: \text{From} \text{Zurek-Kibble} to quantum
  \text{Kibble-Zurek} mechanism},\ }\href
  {https://doi.org/10.1142/S0217751X1430018X} {\bibfield  {journal} {\bibinfo
  {journal} {Int. J. Mod. Phys. A}\ }\textbf {\bibinfo {volume} {29}},\
  \bibinfo {pages} {1430018} (\bibinfo {year} {2014})}\BibitemShut {NoStop}%
\bibitem [{\citenamefont {Dziarmaga}(2010)}]{dziarmaga2010dynamics}%
  \BibitemOpen
  \bibfield  {author} {\bibinfo {author} {\bibfnamefont {J.}~\bibnamefont
  {Dziarmaga}},\ }\bibfield  {title} {\bibinfo {title} {Dynamics of a quantum
  phase transition and relaxation to a steady state},\ }\href
  {https://doi.org/10.1080/00018732.2010.514702} {\bibfield  {journal}
  {\bibinfo  {journal} {Adv. Phys.}\ }\textbf {\bibinfo {volume} {59}},\
  \bibinfo {pages} {1063} (\bibinfo {year} {2010})}\BibitemShut {NoStop}%
\bibitem [{\citenamefont {Kirton}\ \emph {et~al.}(2019)\citenamefont {Kirton},
  \citenamefont {Roses}, \citenamefont {Keeling},\ and\ \citenamefont
  {Dalla~Torre}}]{kirton2019introduction}%
  \BibitemOpen
  \bibfield  {author} {\bibinfo {author} {\bibfnamefont {P.}~\bibnamefont
  {Kirton}}, \bibinfo {author} {\bibfnamefont {M.~M.}\ \bibnamefont {Roses}},
  \bibinfo {author} {\bibfnamefont {J.}~\bibnamefont {Keeling}},\ and\ \bibinfo
  {author} {\bibfnamefont {E.~G.}\ \bibnamefont {Dalla~Torre}},\ }\bibfield
  {title} {\bibinfo {title} {Introduction to the \text{Dicke} model: From
  equilibrium to nonequilibrium, and \emph{vice versa}},\ }\href
  {https://doi.org/10.1002/qute.201800043} {\bibfield  {journal} {\bibinfo
  {journal} {Adv. Quantum Technol.}\ }\textbf {\bibinfo {volume} {2}},\
  \bibinfo {pages} {1800043} (\bibinfo {year} {2019})}\BibitemShut {NoStop}%
\bibitem [{\citenamefont {Keeling}\ \emph {et~al.}(2010)\citenamefont
  {Keeling}, \citenamefont {Bhaseen},\ and\ \citenamefont
  {Simons}}]{keeling2010collective}%
  \BibitemOpen
  \bibfield  {author} {\bibinfo {author} {\bibfnamefont {J.}~\bibnamefont
  {Keeling}}, \bibinfo {author} {\bibfnamefont {M.~J.}\ \bibnamefont
  {Bhaseen}},\ and\ \bibinfo {author} {\bibfnamefont {B.~D.}\ \bibnamefont
  {Simons}},\ }\bibfield  {title} {\bibinfo {title} {Collective dynamics of
  \text{Bose-Einstein} condensates in optical cavities},\ }\href
  {https://doi.org/10.1103/PhysRevLett.105.043001} {\bibfield  {journal}
  {\bibinfo  {journal} {Phys. Rev. Lett.}\ }\textbf {\bibinfo {volume} {105}},\
  \bibinfo {pages} {043001} (\bibinfo {year} {2010})}\BibitemShut {NoStop}%
\bibitem [{\citenamefont {Larson}\ and\ \citenamefont
  {Irish}(2017)}]{larson2017some}%
  \BibitemOpen
  \bibfield  {author} {\bibinfo {author} {\bibfnamefont {J.}~\bibnamefont
  {Larson}}\ and\ \bibinfo {author} {\bibfnamefont {E.~K.}\ \bibnamefont
  {Irish}},\ }\bibfield  {title} {\bibinfo {title} {Some remarks on
  ‘superradiant’ phase transitions in light-matter systems},\ }\href
  {https://doi.org/10.1088/1751-8121/aa65dc} {\bibfield  {journal} {\bibinfo
  {journal} {J. Phys. A: Math. Theor.}\ }\textbf {\bibinfo {volume} {50}},\
  \bibinfo {pages} {174002} (\bibinfo {year} {2017})}\BibitemShut {NoStop}%
\bibitem [{\citenamefont {Soriente}\ \emph {et~al.}(2018)\citenamefont
  {Soriente}, \citenamefont {Donner}, \citenamefont {Chitra},\ and\
  \citenamefont {Zilberberg}}]{soriente2018dissipation}%
  \BibitemOpen
  \bibfield  {author} {\bibinfo {author} {\bibfnamefont {M.}~\bibnamefont
  {Soriente}}, \bibinfo {author} {\bibfnamefont {T.}~\bibnamefont {Donner}},
  \bibinfo {author} {\bibfnamefont {R.}~\bibnamefont {Chitra}},\ and\ \bibinfo
  {author} {\bibfnamefont {O.}~\bibnamefont {Zilberberg}},\ }\bibfield  {title}
  {\bibinfo {title} {Dissipation-induced anomalous multicritical phenomena},\
  }\href {https://doi.org/10.1103/PhysRevLett.120.183603} {\bibfield  {journal}
  {\bibinfo  {journal} {Phys. Rev. Lett.}\ }\textbf {\bibinfo {volume} {120}},\
  \bibinfo {pages} {183603} (\bibinfo {year} {2018})}\BibitemShut {NoStop}%
\bibitem [{\citenamefont {Lu}\ \emph {et~al.}(2023)\citenamefont {Lu},
  \citenamefont {Maiti}, \citenamefont {Garmon}, \citenamefont {Ganjam},
  \citenamefont {Zhang}, \citenamefont {Claes}, \citenamefont {Frunzio},
  \citenamefont {Girvin},\ and\ \citenamefont
  {Schoelkopf}}]{Lu2023ParametricBeamsplitting}%
  \BibitemOpen
  \bibfield  {author} {\bibinfo {author} {\bibfnamefont {Y.}~\bibnamefont
  {Lu}}, \bibinfo {author} {\bibfnamefont {A.}~\bibnamefont {Maiti}}, \bibinfo
  {author} {\bibfnamefont {J.~W.~O.}\ \bibnamefont {Garmon}}, \bibinfo {author}
  {\bibfnamefont {S.}~\bibnamefont {Ganjam}}, \bibinfo {author} {\bibfnamefont
  {Y.}~\bibnamefont {Zhang}}, \bibinfo {author} {\bibfnamefont
  {J.}~\bibnamefont {Claes}}, \bibinfo {author} {\bibfnamefont
  {L.}~\bibnamefont {Frunzio}}, \bibinfo {author} {\bibfnamefont {S.~M.}\
  \bibnamefont {Girvin}},\ and\ \bibinfo {author} {\bibfnamefont {R.~J.}\
  \bibnamefont {Schoelkopf}},\ }\bibfield  {title} {\bibinfo {title}
  {High-fidelity parametric beamsplitting with a parity-protected converter},\
  }\href {https://doi.org/10.1038/s41467-023-41104-0} {\bibfield  {journal}
  {\bibinfo  {journal} {Nat. Commun.}\ }\textbf {\bibinfo {volume} {14}},\
  \bibinfo {pages} {5767} (\bibinfo {year} {2023})}\BibitemShut {NoStop}%
\bibitem [{\citenamefont {Chapman}\ \emph {et~al.}(2023)\citenamefont
  {Chapman}, \citenamefont {de~Graaf}, \citenamefont {Xue}, \citenamefont
  {Zhang}, \citenamefont {Teoh}, \citenamefont {Curtis}, \citenamefont
  {Tsunoda}, \citenamefont {Eickbusch}, \citenamefont {Read}, \citenamefont
  {Koottandavida}, \citenamefont {Mundhada}, \citenamefont {Frunzio},
  \citenamefont {Devoret}, \citenamefont {Girvin},\ and\ \citenamefont
  {Schoelkopf}}]{Chapman2023HighOnOffBeamsplitter}%
  \BibitemOpen
  \bibfield  {author} {\bibinfo {author} {\bibfnamefont {B.~J.}\ \bibnamefont
  {Chapman}}, \bibinfo {author} {\bibfnamefont {S.~J.}\ \bibnamefont
  {de~Graaf}}, \bibinfo {author} {\bibfnamefont {S.~H.}\ \bibnamefont {Xue}},
  \bibinfo {author} {\bibfnamefont {Y.}~\bibnamefont {Zhang}}, \bibinfo
  {author} {\bibfnamefont {J.}~\bibnamefont {Teoh}}, \bibinfo {author}
  {\bibfnamefont {J.~C.}\ \bibnamefont {Curtis}}, \bibinfo {author}
  {\bibfnamefont {T.}~\bibnamefont {Tsunoda}}, \bibinfo {author} {\bibfnamefont
  {A.}~\bibnamefont {Eickbusch}}, \bibinfo {author} {\bibfnamefont {A.~P.}\
  \bibnamefont {Read}}, \bibinfo {author} {\bibfnamefont {A.}~\bibnamefont
  {Koottandavida}}, \bibinfo {author} {\bibfnamefont {S.~O.}\ \bibnamefont
  {Mundhada}}, \bibinfo {author} {\bibfnamefont {L.}~\bibnamefont {Frunzio}},
  \bibinfo {author} {\bibfnamefont {M.~H.}\ \bibnamefont {Devoret}}, \bibinfo
  {author} {\bibfnamefont {S.~M.}\ \bibnamefont {Girvin}},\ and\ \bibinfo
  {author} {\bibfnamefont {R.~J.}\ \bibnamefont {Schoelkopf}},\ }\bibfield
  {title} {\bibinfo {title} {High-on-off-ratio beam-splitter interaction for
  gates on bosonically encoded qubits},\ }\href
  {https://doi.org/10.1103/PRXQuantum.4.020355} {\bibfield  {journal} {\bibinfo
   {journal} {PRX Quantum}\ }\textbf {\bibinfo {volume} {4}},\ \bibinfo {pages}
  {020355} (\bibinfo {year} {2023})}\BibitemShut {NoStop}%
\bibitem [{\citenamefont {Wulschner}\ \emph {et~al.}(2016)\citenamefont
  {Wulschner}, \citenamefont {Goetz}, \citenamefont {Koessel}, \citenamefont
  {Hoffmann}, \citenamefont {Baust}, \citenamefont {Eder}, \citenamefont
  {Fischer}, \citenamefont {Haeberlein}, \citenamefont {Schwarz}, \citenamefont
  {Pernpeintner}, \citenamefont {Xie}, \citenamefont {Zhong}, \citenamefont
  {Zollitsch}, \citenamefont {Peropadre}, \citenamefont {Garcia~Ripoll},
  \citenamefont {Solano}, \citenamefont {Fedorov}, \citenamefont {Menzel},
  \citenamefont {Deppe}, \citenamefont {Marx},\ and\ \citenamefont
  {Gross}}]{Wulschner2016TunableCoupling}%
  \BibitemOpen
  \bibfield  {author} {\bibinfo {author} {\bibfnamefont {F.}~\bibnamefont
  {Wulschner}}, \bibinfo {author} {\bibfnamefont {J.}~\bibnamefont {Goetz}},
  \bibinfo {author} {\bibfnamefont {F.~R.}\ \bibnamefont {Koessel}}, \bibinfo
  {author} {\bibfnamefont {E.}~\bibnamefont {Hoffmann}}, \bibinfo {author}
  {\bibfnamefont {A.}~\bibnamefont {Baust}}, \bibinfo {author} {\bibfnamefont
  {P.}~\bibnamefont {Eder}}, \bibinfo {author} {\bibfnamefont {M.}~\bibnamefont
  {Fischer}}, \bibinfo {author} {\bibfnamefont {M.}~\bibnamefont {Haeberlein}},
  \bibinfo {author} {\bibfnamefont {M.~J.}\ \bibnamefont {Schwarz}}, \bibinfo
  {author} {\bibfnamefont {M.}~\bibnamefont {Pernpeintner}}, \bibinfo {author}
  {\bibfnamefont {E.}~\bibnamefont {Xie}}, \bibinfo {author} {\bibfnamefont
  {L.}~\bibnamefont {Zhong}}, \bibinfo {author} {\bibfnamefont {C.~W.}\
  \bibnamefont {Zollitsch}}, \bibinfo {author} {\bibfnamefont {B.}~\bibnamefont
  {Peropadre}}, \bibinfo {author} {\bibfnamefont {J.-J.}\ \bibnamefont
  {Garcia~Ripoll}}, \bibinfo {author} {\bibfnamefont {E.}~\bibnamefont
  {Solano}}, \bibinfo {author} {\bibfnamefont {K.~G.}\ \bibnamefont {Fedorov}},
  \bibinfo {author} {\bibfnamefont {E.~P.}\ \bibnamefont {Menzel}}, \bibinfo
  {author} {\bibfnamefont {F.}~\bibnamefont {Deppe}}, \bibinfo {author}
  {\bibfnamefont {A.}~\bibnamefont {Marx}},\ and\ \bibinfo {author}
  {\bibfnamefont {R.}~\bibnamefont {Gross}},\ }\bibfield  {title} {\bibinfo
  {title} {Tunable coupling of transmission-line microwave resonators mediated
  by an rf \text{SQUID}},\ }\href
  {https://doi.org/10.1140/epjqt/s40507-016-0048-2} {\bibfield  {journal}
  {\bibinfo  {journal} {EPJ Quantum Technol.}\ }\textbf {\bibinfo {volume}
  {3}},\ \bibinfo {pages} {10} (\bibinfo {year} {2016})}\BibitemShut {NoStop}%
\bibitem [{\citenamefont {Joshi}\ and\ \citenamefont
  {Mahesh}(2022)}]{Joshi2022NMRQuantumBattery}%
  \BibitemOpen
  \bibfield  {author} {\bibinfo {author} {\bibfnamefont {J.}~\bibnamefont
  {Joshi}}\ and\ \bibinfo {author} {\bibfnamefont {T.~S.}\ \bibnamefont
  {Mahesh}},\ }\bibfield  {title} {\bibinfo {title} {Experimental investigation
  of a quantum battery using star-topology \text{NMR} spin systems},\ }\href
  {https://doi.org/10.1103/PhysRevA.106.042601} {\bibfield  {journal} {\bibinfo
   {journal} {Phys. Rev. A}\ }\textbf {\bibinfo {volume} {106}},\ \bibinfo
  {pages} {042601} (\bibinfo {year} {2022})}\BibitemShut {NoStop}%
\end{thebibliography}
\end{document}